\documentclass[twocolumn,trackchanges,twocolappendix]{aastex701}

\newcommand{\pixedfit}{\texttt{piXedfit}}

\shorttitle{Interplay of Compaction, Quenching, and Black Hole Growth}
\shortauthors{Haryana et al.}
\usepackage{xcolor}

\begin{document}

\title{Interplay of Compaction, Quenching, and Black Hole Growth in the Most Massive Galaxies since $z\sim5$: Insights from JWST and Chandra Data}

\author[0009-0009-3404-5673]{Novan Saputra Haryana}
\affiliation{Astronomical Institute, Tohoku University, 6-3, Aramaki, Aoba-ku, Sendai, Miyagi, 980-8578, Japan}
\email[show]{novan.haryana@astr.tohoku.ac.jp}  

\correspondingauthor{Novan Saputra Haryana}

\author[0000-0002-2651-1701]{Masayuki Akiyama}
\affiliation{Astronomical Institute, Tohoku University, 6-3, Aramaki, Aoba-ku, Sendai, Miyagi, 980-8578, Japan}
\email{akiyama@astr.tohoku.ac.jp}

\author[0000-0002-5258-8761]{Abdurro'uf}
\affiliation{Department of Astronomy, Indiana University, Bloomington, IN 47405, USA}
\email{fnuabdurr@iu.edu}

\author[0000-0002-9217-1696]{Suchetha Cooray}
\affiliation{Kavli Institute for Particle Astrophysics and Cosmology, Stanford University, 452 Lomita Mall, Stanford, CA 94305, USA}
\affiliation{SLAC National Accelerator Laboratory, 2575 Sand Hill Road, Menlo Park, CA 94025, USA}
\email{cooray@stanford.edu}

\author[0000-0002-5956-8018]{Itsna Khoirul Fitriana}
\affiliation{Astronomy Research Group, Institut Teknologi Bandung, Jl. Ganesha No. 10 Bandung 40132, Indonesia}
\email{}
\affiliation{National Astronomical Observatory of Japan, 2-21-1, Osawa, Mitaka, Tokyo 181-8588, Japan}

\author[0000-0002-4752-128X]{Dian P. Triani}
\affiliation{Institute for Theory and Computation, Harvard-Smithsonian Center for Astrophysics \\
Cambridge, MA 02138, USA}
\affiliation{Center for Astrophysics \text{\textbar} Harvard \& Smithsonian, 60 Garden Street, Cambridge, MA 02138, USA}
\email{}

\author[0000-0003-2213-7983]{Bovornpratch Vijarnwannaluk}
\affiliation{Academia Sinica Institute of Astronomy and Astrophysics (ASIAA), 11F of Astronomy-Mathematics Building, AS/NTU, No. 1, Section 4, 12 Roosevelt Road,
Taipei 106319, Taiwan}
\email{}

\author[0009-0007-3423-1332]{Juan Pablo Alfonzo}
\affiliation{Astronomical Institute, Tohoku University, 6-3, Aramaki, Aoba-ku, Sendai, Miyagi, 980-8578, Japan}
\email{}

\author[0000-0001-6102-9526]{Irham Taufik Andika}
\affiliation{Technical University of Munich, TUM School of Natural Sciences, Department of Physics, James-Franck-Str. 1, D-85748 Garching,
Germany}
\affiliation{Max-Planck-Institut für Astrophysik, Karl-Schwarzschild-Str. 1, D-85748 Garching, Germany}
\email{}

\author[0009-0006-8034-6061]{Muhammad Nur Ihsan Effendi}
\affiliation{Master’s Program in Astronomy, Institut Teknologi Bandung, Jl. Ganesha No. 10 Bandung 40132, Indonesia}
\email{}

\author[0000-0002-5897-2158]{Ibnu Nurul Huda}
\affiliation{School of Astronomy and Space Science, Key Laboratory of Modern Astronomy and Astrophysics (Ministry of Education), Nanjing University, No. 163 Xianlin Avenue, 210023 Nanjing, People's Republic of China}
\affiliation{Research Center for Computation, National Research and Innovation Agency, Jl. Sangkuriang No. 10, 40135 Bandung, Indonesia}
\email{}

\author[0000-0003-2918-9890]{Takumi Kakimoto}
\affiliation{Department of Astronomical Science, The Graduate University for Advanced Studies, SOKENDAI, 2-21-1 Osawa, Mitaka, Tokyo 181-8588, Japan}
\affiliation{National Astronomical Observatory of Japan, 2-21-1, Osawa, Mitaka, Tokyo 181-8588, Japan}
\email{}

\author[0000-0001-6282-5778]{Anton Timur Jaelani}
\affiliation{Astronomy Research Group, Institut Teknologi Bandung, Jl. Ganesha No. 10 Bandung 40132, Indonesia}
\email{}

\author[0000-0002-0322-6131]{Ronaldo Laishram}
\affiliation{National Astronomical Observatory of Japan, 2-21-1, Osawa, Mitaka, Tokyo 181-8588, Japan}
\email{}

\author[0000-0002-8299-0006]{Naoki Matsumoto}
\affiliation{Astronomical Institute, Tohoku University, 6-3, Aramaki, Aoba-ku, Sendai, Miyagi, 980-8578, Japan}
\email{}

\author[0000-0001-7713-0434]{Abdurrahman Naufal}
\affiliation{Academia Sinica Institute of Astronomy and Astrophysics (ASIAA), 11F of Astronomy-Mathematics Building, AS/NTU, No. 1, Section 4, 12 Roosevelt Road,
Taipei 106319, Taiwan}
\email{}

\author[0000-0002-6386-5373]{Lucky Puspitarini}
\affiliation{Astronomy Research Group, Institut Teknologi Bandung, Jl. Ganesha No. 10 Bandung 40132, Indonesia}
\email{}

\author[0009-0005-1487-7772]{Ryo Albert Sutanto}
\affiliation{Astronomical Institute, Tohoku University, 6-3, Aramaki, Aoba-ku, Sendai, Miyagi, 980-8578, Japan}
\email{}

\author[0000-0003-1551-519X]{Hesti Retno Tri Wulandari}
\affiliation{Astronomy Research Group, Institut Teknologi Bandung, Jl. Ganesha No. 10 Bandung 40132, Indonesia}
\email{}

\begin{abstract}

The buildup of dense stellar cores is expected to mark an important transition in the star-formation and black-hole growth of massive galaxies. Using spatially resolved spectral energy distribution (SED) fitting of James Webb Space Telescope near-infrared imaging, combined with stacking analysis of Chandra X-ray data, we trace stellar mass buildup and average black hole accretion in the most massive galaxies at $z<5$, selecting 50 most massive galaxies per redshift bin at constant number density of {{$\sim4.4\times10^{-5}$ cMpc$^{-3}$}}. To robustly constrain central stellar populations, we separate active galactic nuclei (AGN) components affecting the photometry using multi-band morphological decomposition and SED analysis. {We find that the sample selected with constant number density exhibits evolutionary trend of rapid central compaction at $z\sim4$}, during which the median central 1 kpc stellar mass increases by {$\sim0.60$} dex over $\sim400$ Myr. The majority of X-ray detected AGN ({$63\%\pm12\%$}) are hosted by galaxies undergoing the compaction, {while we find neither individually detected X-ray sources nor a significant stacked X-ray signal at $z>4$, indicating that substantial average black-hole growth emerges primarily during, rather than before, the compaction}. Following the compaction, central specific star formation rates (sSFR) decline {by {{$\sim1.24$}} dex over $\sim700$ Myr} at $z\sim3$ while remaining elevated galaxy-wide, signaling the onset of inside-out quenching. Despite this central suppression, specific black hole accretion rate remains coupled to the total sSFR. {Our results suggest that dense-core formation in the most massive galaxies marks the onset of inside-out quenching and a transition toward enhanced black-hole to stellar growth ratio.}

\end{abstract}


\section{Introduction}

The formation and quenching of massive galaxies are widely recognized as key open questions in galaxy evolution, yet the physical mechanisms governing these transitions remain poorly understood. In particular, the interplay between central stellar mass assembly, the suppression of star formation, and the growth of supermassive black holes (SMBHs) continues to be actively debated, especially at high redshift where galaxies undergo their most rapid structural and baryonic evolution.

Both observations and simulations have shown that many massive galaxies experience a phase of rapid central stellar mass buildup, often referred to as compaction \citep[e.g.,][]{2015MNRAS.450.2327Z, 2016MNRAS.457.2790T, 2017ApJ...840...47B}. During this phase, galaxies develop dense stellar cores with surface mass densities comparable to those of present-day early-type galaxies \citep[e.g.,][]{2025ApJ...994..215H}, even while maintaining high star formation rates. Compaction is thought to be driven by a variety of dissipative processes, including gas-rich inflows, mergers, violent disk instabilities, and clump migration, but their relative importance remains uncertain \citep{2015MNRAS.450.2327Z}.

Following compaction, many galaxies exhibit a decline in star formation that proceeds from the central regions outward, commonly described as inside-out quenching \citep{2017ApJ...840...47B, 2025ApJ...994..215H, 2026ApJ...998..158L}. Several physical mechanisms have been proposed to explain this transition. In some scenarios, quenching is linked to gas depletion in the central regions when the inflow rate falls below the combined rate of star formation and outflows \citep[][]{2010ApJ...718.1001B, 2013ApJ...772..119L, 2021Natur.597..485W}. In others, the buildup of a dense stellar core stabilizes the remaining gas against fragmentation, reducing star formation efficiency without necessarily removing the gas \citep{2009ApJ...707..250M}. Feedback from active galactic nuclei (AGN) may further contribute by heating or expelling gas, though its role relative to gravitational and stellar processes is still under investigation \citep[e.g.,][]{2005Natur.433..604D, 2012ARA&A..50..455F, 2025MNRAS.543.1878L, 2025MNRAS.539.3568A, 2026arXiv260214496H}.

At the same time, the co-evolution of SMBHs and their host galaxies introduces additional complexity. Empirical correlations between black hole mass and galaxy properties (e.g., stellar mass, velocity dispersion, etc.) suggest a close connection between black hole growth and stellar mass assembly \citep{2013ARA&A..51..511K, 2012ApJ...753L..30M, 2020A&A...642A..65C}. However, it remains unclear how this coupling evolves across different phases of galaxy evolution, particularly during and after central quenching. Recent theoretical work suggests that black hole growth may proceed through multiple regimes, ranging from star-formation–dominated growth at early times to feedback-regulated accretion at later stages \citep{2025MNRAS.543.1878L}. Whether and how black hole accretion continues once central star formation is suppressed remains an open observational question.

{Addressing these issues requires spatially resolved measurements of stellar mass and star formation in high-redshift galaxies, to separate the central and total properties, combined with robust tracers of black-hole accretion. The advent of the James Webb Space Telescope (JWST) now enables resolved studies of galaxy structure and star formation at $z\sim5$ \citep[e.g.,][]{2024ApJ...974..135J,2025MNRAS.539.2685L,2025ApJ...994..215H, Laishram_et_al_2026c}. At the same time, its high angular resolution and rest-frame optical coverage provide new opportunities to separate AGN and host-galaxy light, which is essential for interpreting the stellar structure of active galaxies.}

{Recent JWST studies have begun to characterize the host galaxies of high-redshift AGN through multi-band morphological decomposition. For example, \citet{2024ApJ...962...93Z} used COSMOS-Web NIRCam imaging to decompose X-ray-selected broad-line AGN and their host galaxies, demonstrating that host-galaxy properties can be recovered even in the presence of unresolved nuclear emission. Similarly, \citet{2025ApJ...994..265V} studied the stellar morphology and size--mass relation of X-ray-selected AGN hosts with JWST, showing that AGN hosts occupy structural regimes between star-forming and quiescent galaxies. These studies highlight the importance of treating the unresolved AGN component when measuring host-galaxy structure, stellar mass, and star formation.
}

{However, most existing JWST AGN-host studies are based on AGN-selected samples and therefore primarily address the properties of galaxies once AGN activity is already identified. A complementary approach is to start from a mass-selected galaxy population and ask when, during the structural evolution of massive galaxies, black-hole accretion becomes prominent. This is particularly important for testing whether black-hole growth is connected to the formation of dense stellar cores and the subsequent onset of central star-formation suppression.
}

{In this paper, we combine spatially resolved JWST near-infrared imaging with Chandra X-ray data, using both individual detections and stacking analysis, to investigate the evolution of central stellar mass density, star formation, and black-hole accretion in the most massive galaxies at $z<5$. Our sample is selected to trace the likely progenitors of present-day massive systems by selecting the most massive galaxies in redshift bins of approximately equal comoving volumes. To robustly evaluate the stellar populations in the central regions, we separate AGN components that affect the photometry using multi-band morphological decomposition and spectral energy distribution analysis. By comparing the evolution of central and total star formation with X-ray-inferred black-hole accretion, we aim to place new observational constraints on how compaction, inside-out quenching, and SMBH growth are connected in massive galaxies. We assume a $\Lambda$CDM model with the cosmological parameters $H_0=70$ km s$^{-1}$ Mpc$^{-1}$, $\Omega_M = 0.3$, and $\Omega_\Lambda=0.7$ throughout this paper.}

\section{Data and Sample Selection}
\subsection{JWST and HST Imaging Data}

We use publicly available JWST imaging data from three major extragalactic legacy fields: the Extended Groth Strip (EGS; ceers-full-grizli-v7.2), COSMOS (divided into primer-cosmos-west-grizli-v7.0 and primer-cosmos-east-grizli-v7.0), and the Ultra Deep Survey (UDS; divided into primer-uds-south-grizli-v7.2 and primer-uds-north-grizli-v7.2). These fields have been extensively covered by the Hubble Space Telescope (HST) as part of the Cosmic Assembly Near-infrared Deep Extragalactic Legacy Survey (CANDELS; \citealt{2011ApJS..197...35G}, \citealt{2011ApJS..197...36K}), ensuring a robust set of complementary optical to near-infrared observations.

The JWST imaging data are taken from the DAWN JWST Archive (DJA)\footnote{\url{https://dawn-cph.github.io/dja/imaging/v7/}}, v7 version of the reductions. These data were processed using the \texttt{GRIZLI} package \citep{2023zndo...8370018B}, which drizzles them onto a common pixel grid with a pixel scale of $0.^{\prime\prime}04$.  
For a detailed description of the data reduction procedures, please refer to \cite{2023ApJ...947...20V}.

We compile JWST data from several public programs. In the EGS field, we utilize data from the Cosmic Evolution Early Release Science Survey (CEERS; ERS 1345, PI Finkelstein; \citealt{2023ApJ...946L..13F}), combined with additional F444W imaging from GO 2279 (PI Naidu) and data from several other filters from GO 2750 (PI Arrabal-Haro). In total, we use JWST/NIRCam imaging data in seven filters, consisting of F115W, F150W, F200W, F277W, F356W, F410M, and F444W. For the UDS and COSMOS fields, the imaging data were taken from the Public Release Imaging for Extragalactic Research (PRIMER; GO 1837, PI Dunlop), which partially overlaps with the COSMOS-Web survey (GO 1727, PIs Kartaltepe \& Casey; \citealt{2023ApJ...954...31C}). In these fields, we make use of NIRCam data in eight filters, similar to those in EGS, but with the addition of the F090W band.

To supplement these JWST observations, we incorporate archival HST imaging data from the CANDELS survey. For all fields, we include HST/ACS images in the F435W, F606W, and F814W bands. These extensive multi-wavelength datasets are essential for robustly constraining stellar population properties via SED fitting.

DJA also provides photometry catalogs along with basic global properties and redshift measurements derived from SED fitting. The photometry was performed using Source Extractor Python (\texttt{SEP}; \citealt{2016JOSS....1...58B}) and \texttt{GRIZLI}, while the SED fitting was performed with \texttt{eazy-py} \citep{2008ApJ...686.1503B, 2021zndo...7575984B} using the \texttt{agn\_blue\_sfhz\_13}\footnote{\url{https://github.com/gbrammer/eazy-photoz/blob/master/templates/sfhz/README.md}} template set. This template set combines 13 redshift-dependent galaxy templates with additional blue emission-line and AGN-like components to improve photometric-redshift estimates for sources with unusual or AGN-contaminated SEDs. The SED fitting incorporates additional MIRI JWST photometry. In addition to photometric redshift measurements, the catalogs also provide spectroscopic redshifts compiled from various sources in the literature. 

To assess the reliability of the photometric redshifts, we compare predicted photometric redshifts with available spectroscopic measurements. We obtain that \( |z_{\text{spec}} - z_{\text{phot}}| / (1 + z_{\text{spec}}) \) has a median of \( 1.45 \times 10^{-2} \) with a normalized median absolute deviation (NMAD) of \( \sigma_{\text{NMAD}} \simeq 0.017 \). We use \( \sigma_{\text{NMAD}} \) as \citep[following][]{2008ApJ...686.1503B}  
\begin{equation}
    \sigma_{\text{NMAD}} = 1.48 \times \text{median} \left( \frac{|\Delta z - \text{median}(\Delta z)|}{1 + z_{\text{spec}}} \right),
\end{equation}  
where \( \Delta z = |z_{\text{spec}} - z_{\text{phot}}| \). Only 7\% of galaxies meet the outlier criterion of \( |z_{\text{spec}} - z_{\text{phot}}| / (1 + z_{\text{spec}}) > 0.15 \). These metrics indicate that the photometric redshifts are reliable.

\subsection{Sample Selection}

In this subsection, we describe how we construct a sample of massive galaxies across $0 < z < 5$ that is intended to trace possible evolutionary connections. We first select all galaxies with $z < 5$ and $M_\ast > 3\times10^{9}M_\odot$ based on photometric redshifts (or spectroscopic redshifts when available) and stellar masses from the \texttt{eazy-py} catalog, yielding an initial sample of 19447 galaxies. We then remove artifacts such as stellar contamination, sources close to bright stars, or objects located near the edges of the field—through visual inspection. We also remove all galaxies that are fully covered in fewer than 6 filters, resulting in a sample of 19204 galaxies. For the remaining galaxies, we refit the integrated photometry using \texttt{piXedfit} (see Section~\ref{sec:SED_fitting} for details of the fitting setup) in order to obtain more accurate estimates of the total stellar mass used for our final selection.

We divide the resulting sample into five redshift bins with approximately equal comoving volumes. To construct a sample that can plausibly represent an evolutionary sequence, we select the most massive galaxies in each redshift bin, such that they follow a constant comoving number density across redshift \citep[e.g.,][]{2010ApJ...709.1018V, 2025ApJ...994..215H}. This selection assumes that the most massive galaxies at high redshift evolve into the most massive galaxies in the local universe \citep[e.g.,][]{2013ApJ...766...33L}. We adopt a fixed number of 50 galaxies per bin, corresponding to the number of galaxies available in the highest redshift bin, resulting in a constant number density of {$\sim4.4\times10^{-5}\mathrm{cMpc}^{-3}$}. {We acknowledge that this assumption is not exact, because mergers and stochastic growth histories can change the rank ordering of galaxies with time. We discuss this caveat in Section \ref{subsec}, where we test the effect of adopting an evolving number-density selection following \cite{Behroozi2013}.}

Galaxies with substantial AGN contamination can have overestimated stellar masses and inaccurate redshift estimates, and may therefore not genuinely belong to our sample. We thus exclude these galaxies from the analysis. We replace them with the next most massive galaxies, iterating until no contaminated sources remain. Such sources are identified by (i) $ f_{\mathrm{F444W}} > 0.5$ and (ii) $\Delta \mathrm{AIC} > 0$, where $f_{\mathrm{F444W}}$ is the fraction of F444W flux in the unresolved nuclear component relative to the total F444W flux, and $\Delta \mathrm{AIC}$ indicates whether the nuclear component is more likely due to AGN emission rather than unresolved stellar light. These metrics are described in Section~\ref{subsec:byebyeAGN}. In total, we identify four such objects: two at $z \sim 3.1$ (with inferred $\log(M_*/M_\odot) = 10.9$ and 10.7) and two at $z \sim 0.9$ (with inferred $\log(M_*/M_\odot) = 11.0$ and 10.6).

\begin{figure}
    \centering
    \includegraphics[width=\linewidth]{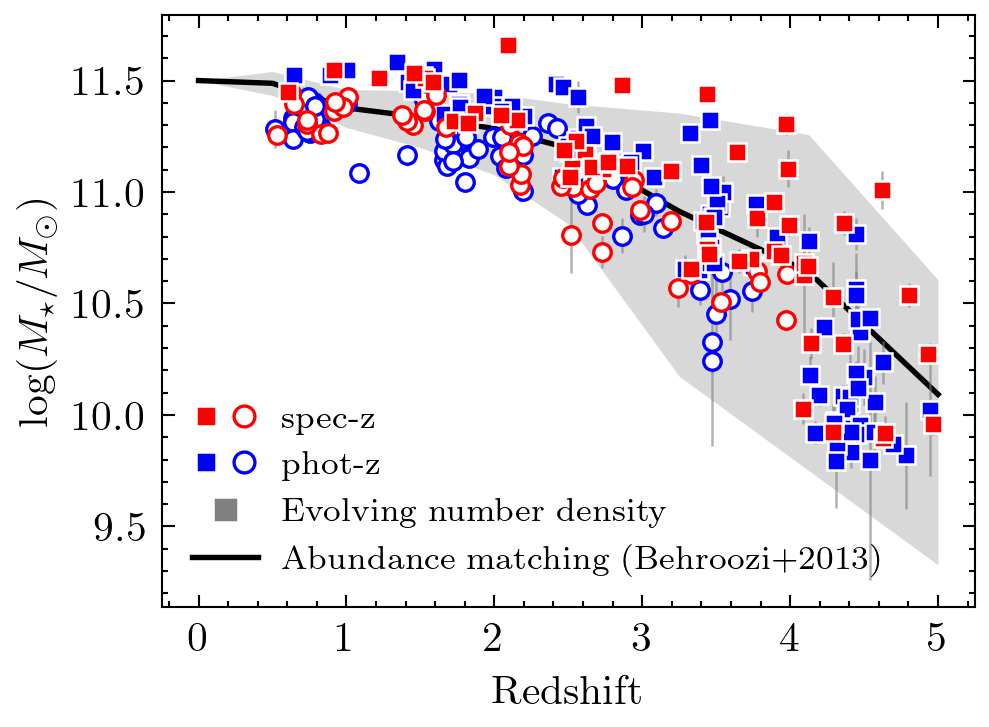}
    \caption{Stellar mass as a function of redshift for our sample. Blue points indicate galaxies with photometric redshifts, while red points denote galaxies with confirmed spectroscopic redshifts. Square symbols indicate galaxies selected using the evolving number-density criterion of \citet{Behroozi2013}, which is discussed further in Section~\ref{subsec}. The black solid line shows the expected stellar mass evolution of galaxies with $\log(M_*/M_\odot)=11.5$ at $z=0$, estimated using the \texttt{nd-redshift} code \citep{Behroozi2013}. The grey shaded area is showing the $1\sigma$ uncertainty of the \texttt{nd-redshift} prediction. The stellar masses are obtained from spatially resolved SED fitting (Section \ref{sec:SED_fitting}) after performing AGN removal (Section \ref{subsec:byebyeAGN}).}
    \label{fig:massevol}
\end{figure}

We summarize the redshift ranges, stellar mass, the number of X-ray detected galaxies (explained in the next subsection), and the number of spectroscopically confirmed galaxies in each bin in Table~\ref{tab:numgal}. In total, we have {98 galaxies} (out of 250) with confirmed spectroscopic redshifts. The stellar mass–redshift distribution of the sample is shown in Figure~\ref{fig:massevol}, where red points indicate galaxies with spectroscopic redshifts. The black solid line shows the expected median stellar mass evolution of galaxies with $\log(M_*/M_\odot)=11.5$ at $z=0$, estimated using the \texttt{nd-redshift} code \citep{Behroozi2013}, which is based on abundance-matching techniques (see \citealt{2025ApJ...994..215H}, Section~4.3 for details). {Square symbols indicate galaxies selected using the evolving number-density criterion of \citet{Behroozi2013}, which is discussed further in Section~\ref{subsec}.}

\begin{table*}
    \centering
    \caption{{Redshift range, median redshift, lowest stellar mass, median stellar mass, number of X-ray–detected galaxies, and number of spectroscopically confirmed galaxies in each bin. Each redshift bin contains 50 galaxies, corresponding to a constant comoving number density of $4.4\times10^{-5}\mathrm{cMpc}^{-3}$.}}
    \label{tab:numgal}
    \begin{tabular}{l l l l l l}
    \hline
        Redshift & Redshift & Lowest & Median & \#X-Ray Det. & \#Spec-z \\
        Range & Median & $\log(M_*/M_\odot)$ & $\log(M_*/M_\odot)$ & Galaxies & Galaxies \\
    \hline
        $0.5 < z < 1.7$ & 0.92$_{-0.27}^{+0.60}$ & 11.08 & $11.37_{-0.09}^{+0.14}$ & 10/50 & 23/50 \\
        $1.7 < z < 2.5$ & 1.99$_{-0.27}^{+0.19}$ & 11.00 & $11.26_{-0.12}^{+0.12}$ & 7/50 & 14/50 \\
        $2.5 < z < 3.2$ & 2.66$_{-0.14}^{+0.33}$ & 10.73 & $11.07_{-0.15}^{+0.14}$ & 9/50 & 24/50 \\
        $3.2 < z < 4.1$ & 3.49$_{-0.09}^{+0.40}$ & 10.24 & $10.74_{-0.17}^{+0.26}$ & 4/50 & 22/50 \\
        $4.1 < z < 5.0$ & 4.41$_{-0.22}^{+0.22}$ & 9.79 & $10.08_{-0.17}^{+0.46}$ & 0/50 & 15/50 \\
        \hline
    \end{tabular}
\end{table*}

\subsection{AGN Catalog Cross-Matching and X-ray Source Selection}
\label{subsec:QSOmatch}

One of the goals of this paper is to examine the connection between the compaction and active galactic nuclei activity. To this end, we cross-match our galaxy sample with publicly available X-ray and quasar catalogs to identify potential AGN hosts.

For the X-ray catalogs, we use Chandra-based surveys in each field: the AEGIS-X Deep survey for EGS \citep{2015ApJS..220...10N}, the X-UDS survey for UDS \citep{2018ApJS..236...48K}, and the COSMOS-Legacy survey for COSMOS \citep{2016ApJ...819...62C}. Objects are classified as X-ray sources if they are detected in any of the soft (0.5–2.0 keV), hard (2.0–10.0 keV for UDS and 2.0–7.0 keV for the other fields), or full band (0.5-10.0 keV for UDS and 0.5-7.0 keV for the other fields). The detection is based on poisson false-probability of $p<1\times10^{-4}$ (corresponding to $\sim3.7\sigma$) for X-UDS, $p<2\times10^{-5}$ ($\sim4.3\sigma$) in COSMOS-Legacy, and $p<4\times10^{-6}$ ($\sim4.5\sigma$) in AEGIS-X 
. All detected sources in our sample have absorption corrected hard-band X-ray intrinsic luminosities of  $L_{\mathrm{2-10keV}}>10^{42}$ erg s$^{-1}$. In addition, we cross-match our sample with spectroscopically identified quasar catalogs, including the SDSS DR16 Quasar Catalog \citep{2020ApJS..250....8L} and the Million Quasar Catalog \citep{2023OJAp....6E..49F}.

All cross-matches are performed using a matching radius of 1 arcsecond (25 NIRCam pixels). Galaxies with counterparts in any of these catalogs are flagged as AGN or QSO candidates. In total, {39} out of 250 galaxies in our sample are classified as possible AGN/QSO hosts, with {32} of them classified with X-ray detection. We show the number of objects with X-ray detection in each redshift bin on the third column of Table \ref{tab:numgal}.

\section{Methods}

\subsection{Empirical Point Spread Function}

We construct empirical point spread function (PSF) using \texttt{PSFEx} \citep[][]{2011ASPC..442..435B} in combination with \texttt{SExtractor} \citep[][]{1996A&AS..117..393B}, following exactly the procedure described in \cite{2025ApJ...994..215H}. A brief summary is provided here; full details are provided in Section 3.1 of that work. First, point sources are identified in the F200W image using the \texttt{MU\_MAX}–\texttt{MAG\_AUTO} diagnostic diagram, where stars form a tight linear locus distinct from extended galaxies. We isolate this stellar locus using a combination of density-based clustering (DBSCAN, \cite{1996kddm.conf..226E}) and robust linear regression (RANSAC, \cite{10.1145/358669.358692}), and select unsaturated point sources within a narrow band around the fitted star sequence. These stars are then used to construct empirical PSFs with \texttt{PSFEx}. The same stars selected in F200W are applied to all other bands to ensure consistent sampling across filters.

\subsection{Identification and Removal of Unresolved AGN Emission}
\label{subsec:byebyeAGN}

To obtain accurate measurements of stellar population properties, it is essential to account for possible contamination from unresolved emission associated with AGN. Here, we describe how we identify and remove AGN contributions in our galaxy images. Our goal is to subtract nuclear point sources from the multi-band imaging in a manner that is consistent with a quasar spectral energy distribution.

We first estimate the flux of any central point source in each band by fitting the multi-band images of each galaxy with \texttt{GALIGHT} \citep{2020ApJ...888...37D}, adopting a model consisting of a single Sérsic profile plus a central PSF component. The Sérsic index, ellipticity, and position angle are fixed to values measured in the band closest to rest-frame 7000\AA\, ensuring a consistent structural model across filters. The fitted PSF component's flux in each band is then used to construct the SED of the unresolved nuclear source.

We next fit the extracted point-source SED with quasar templates. We adopt the SDSS composite quasar spectrum from \cite{2001AJ....122..549V} combined with the near-infrared extension from \cite{2006ApJ...640..579G}, and generate a library of attenuated spectra assuming an SMC extinction curve \citep{2003ApJ...594..279G}. Both the visual extinction and the overall normalization are varied on a grid of 100 values each, yielding $10^4$ templates per galaxy, with a flat prior on $E(B-V)\in[0.0,2.0]$ and a flat prior on the normalization. The normalization is anchored to the measured point-source SED. We compute the posterior probability distribution using a Student-$t$ likelihood and adopt the posterior median as the best-fit quasar model.

However, unresolved central stellar population can also contribute to the nuclear SED. To distinguish between stellar and AGN origins, we also fit the point-source SED with galaxy template libraries using \texttt{piXedfit}, with the same parameter setup described in Section~\ref{sec:SED_fitting}. We compare the quasar and galaxy fits using the Akaike Information Criterion (AIC),
\begin{equation}
    \text{AIC} = 2k - 2\ln {L},
    \label{eq:deltaAIC}
\end{equation}
where $k$ is the number of free parameters and ${L}$ is the maximum likelihood. A smaller AIC indicates a preferred model. We classify a source as AGN-dominated and therefore requiring point-source subtraction if $\Delta \mathrm{AIC}=\text{AIC}_{\rm galaxy} - \text{AIC}_{\rm quasar}>0$ or if the galaxy is flagged as a possible QSO/AGN host (see Section \ref{subsec:QSOmatch}).

Quasars are known to follow a tight, approximately linear relation between their monochromatic 2 keV and 2500\AA\ luminosities \citep[e.g.,][]{2010A&A...512A..34L}. To validate both the quasar SED fitting and the $\Delta\mathrm{AIC}$ classification, we compare the offset of the X-ray luminosity from this established relation with the $\Delta$AIC values. For this purpose, we derive the rest-frame monochromatic 2 keV luminosities (or upper limits for non-detections) from the absorption-corrected 2--10 keV fluxes obtained in Section~\ref{sec:BHAR_estimation}, assuming a power-law spectrum with a photon index of $\Gamma=1.8$. We compare these luminosities with the dust-corrected rest-frame 2500\AA\ luminosities derived from the best-fit quasar templates.

\begin{figure}
    \centering
    \includegraphics[width=\linewidth]{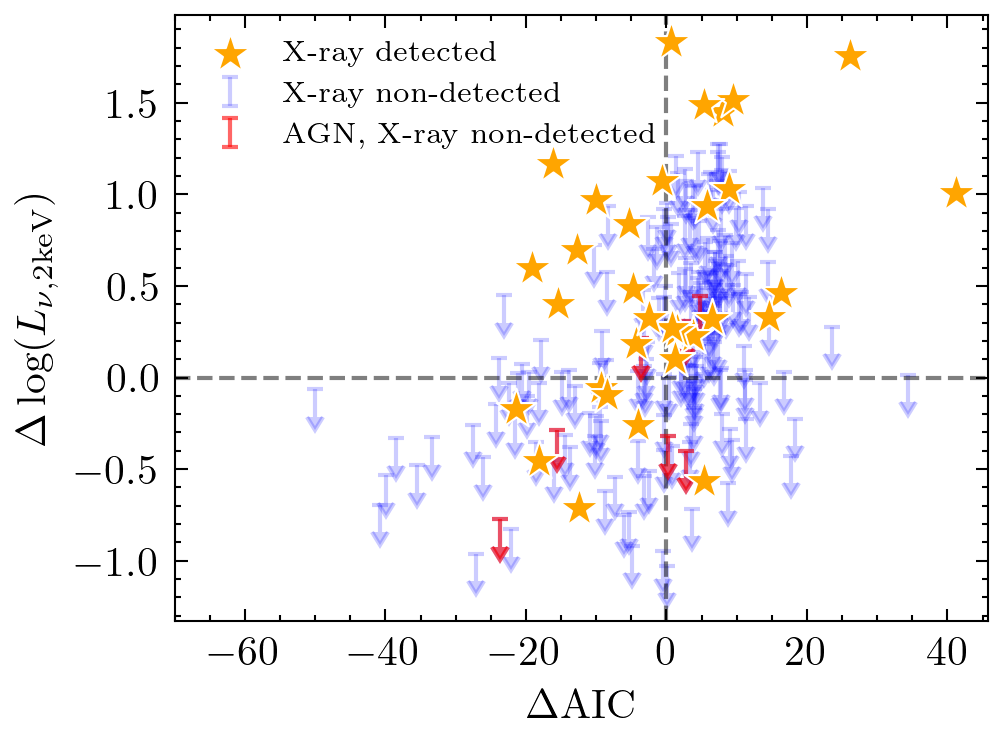}
    \caption{Offset $\Delta \log(L_{\nu,2\mathrm{keV}}) \equiv \log(L_{\nu,2\mathrm{keV}}) - f(L_{\nu,2500\text{\AA}})$ versus $\Delta$AIC, where $f(L_{\nu,2500\text{\AA}})$ is the expected logarithm of 2 keV luminosity from the $L_{\nu,2\mathrm{keV}}$–$L_{\nu,2500\text{\AA}}$ relation of \cite{2010A&A...512A..34L}, evaluated using the quasar SED–derived dust-corrected $L_{\nu,2500\text{\AA}}$. Orange stars denote X-ray detections, while blue and red downward arrows indicate upper limits for non-detections and AGN without X-ray detections, respectively.}
    \label{fig:DeltaL_AIC}
\end{figure}
In Figure \ref{fig:DeltaL_AIC}, we plot the offset
$\Delta \log(L_{\nu,2\mathrm{keV}}) \equiv \log(L_{\nu,2\mathrm{keV}}) - f(L_{\nu,2500\text{\AA}})$
against $\Delta$AIC, where $f(L_{\nu,2500\text{\AA}})$ is the expected logarithm of 2 keV luminosity from the $L_{\nu,2\mathrm{keV}}$–$L_{\nu,2500\text{\AA}}$ relation of \cite{2010A&A...512A..34L}, evaluated using the quasar SED–derived $L_{\nu,2500\text{\AA}}$. X-ray–detected sources (orange stars) lie close to the relation and show a clear trend: larger $\Delta$AIC corresponds to more positive $\Delta \log(L_{\nu,2\mathrm{keV}})$. For X-ray non-detections, we plot $\Delta \log(L_{\nu,2\mathrm{keV}})$ derived from upper limits on $L_{\nu,2\mathrm{keV}}$ (blue downward arrows). These limits are broadly consistent with the relation and follow the same trend, with higher $\Delta$AIC corresponding to higher $\Delta \log(L_{\nu,2\mathrm{keV}})$ upper limits. Quantitatively, {85\%} of galaxies with $\Delta\mathrm{AIC} > 0$ have $\Delta \log(L_{\nu,2\mathrm{keV}}) > 0$, while {55\%} of those with $\Delta\mathrm{AIC} < 0$ have $\Delta \log(L_{\nu,2\mathrm{keV}}) < 0$. This indicates that galaxies with $\Delta\mathrm{AIC} < 0$ are preferentially offset below the \cite{2010A&A...512A..34L} relation, whereas those with $\Delta\mathrm{AIC} > 0$ remain consistent with it, as their upper limits typically lie above the relation. AGN candidates without X-ray detections (red circles with downward arrows) show the same behavior, with $\Delta \log(L_{\nu,2\mathrm{keV}})$ upper limits that also correlate with $\Delta$AIC. Overall, the nuclear point-source flux traces the intrinsic quasar emission, following the established UV–X-ray relation. Moreover, $\Delta$AIC is physically meaningful, correlating with the offset of the X-ray luminosity from the expected relation.

\begin{figure*}
    \centering
    \includegraphics[width=\linewidth]{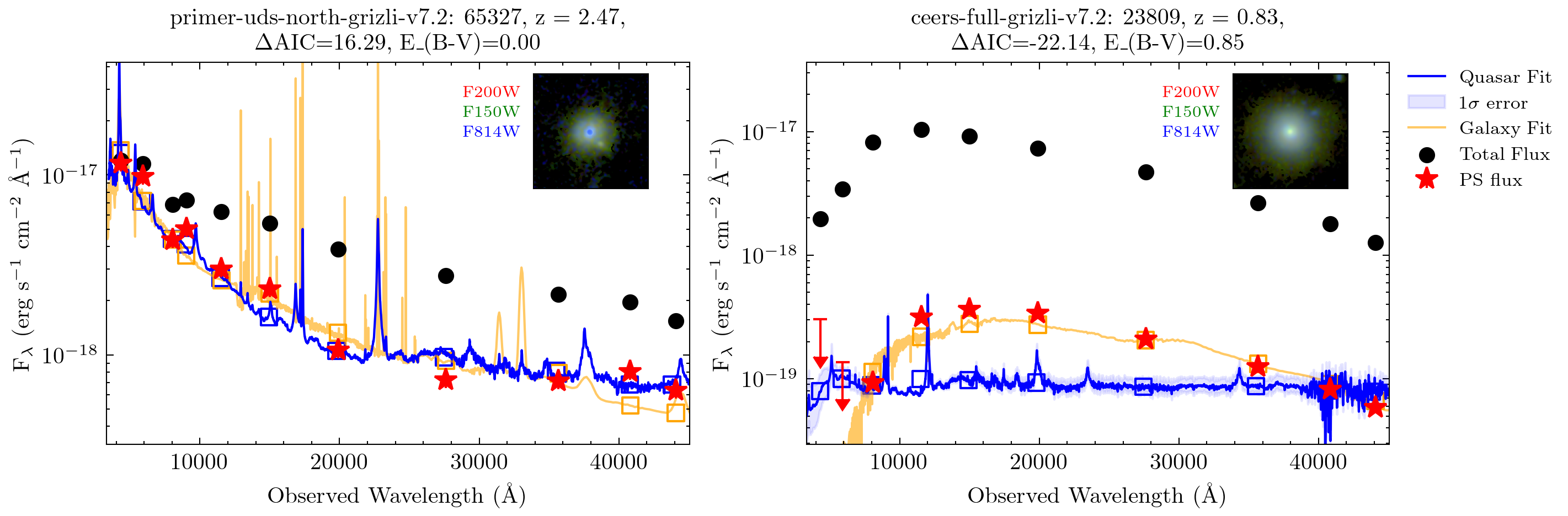}
    \caption{The total SEDs (black circles), point-source SEDs (red stars, from Sérsic+PSF decomposition), and the best-fit quasar and galaxy templates (blue and orange curves, respectively) are shown. RGB images of each galaxy, constructed from PSF-matched F115W, F200W, and F444W images (all matched to the F444W PSF), are displayed in the upper-right corner of each panel. The left panel shows a galaxy with nuclear component better fit by the quasar template, while the right panel better fit by the galaxy template.}
    \label{fig:PS_SED_check}
\end{figure*}

\begin{figure*}
    \centering
    \includegraphics[width=\linewidth]{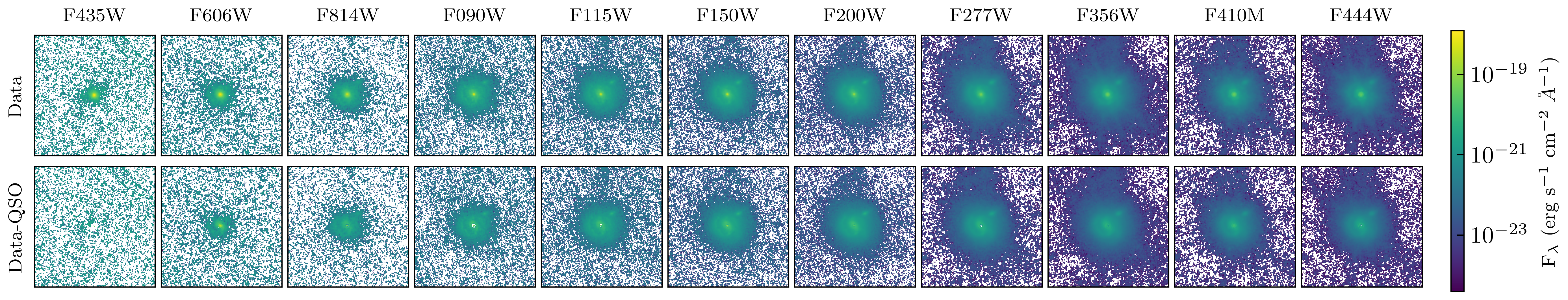}
    \caption{Top: Observed multi-band images of primer-uds-north-grizli-v7.2\#65327. Bottom: The same images after subtraction of a PSF component scaled by the best-fit quasar SED (see Section~\ref{subsec:byebyeAGN} and Figure~\ref{fig:PS_SED_check}).}
    \label{fig:subtracted}
\end{figure*}

For galaxies satisfying the $\Delta$AIC criteria or flagged as a possible QSO/AGN host ({150 sources} in total), we then subtract the nuclear component by scaling the PSF image in each band according to the best-fit quasar SED, shifting it to the band-specific centroid determined by \texttt{GALIGHT}, and removing it from the original image. The resulting images are thus corrected for unresolved AGN components. In Figure \ref{fig:PS_SED_check}, we show example point-source SEDs (red points) together with the total SEDs (black points), best-fit quasar and galaxy templates (blue and orange, respectively) for a galaxy requiring point-source subtraction (left) and a galaxy that does not require subtraction (right). In Figure~\ref{fig:subtracted}, we show an example of the observed images after subtraction of the PSF scaled by the best-fit quasar SED (lower panel).

{Because our subsequent analysis relies on spatially resolved stellar mass and SFR measurements, the accuracy of the nuclear subtraction is particularly important in the central regions. We therefore further assess the reliability of this procedure in Appendix \ref{app: AGN_rec}, where we quantify how the AGN-subtraction method affects the inferred central stellar mass and star formation rate measurements.}

\subsection{Spatially Resolved SED Fitting}

\subsubsection{PSF Matching}

Because our multi-band images span a wide wavelength range, we must match the PSF across all bands to obtain accurate spatial measurements of stellar population properties. To this end, we construct PSF-matching kernels from the PSFs described in Section 3.1 and apply them to the galaxy images.

We perform PSF matching by degrading all images to the spatial resolution of the band with the largest PSF, which in our case is the NIRCam F444W filter. For each filter, we construct convolution kernel using the ratio of its PSF to F444W's PSF in Fourier space. These kernels are designed to transform the PSF of each band into that of F444W. The accuracy of this PSF matching is demonstrated in Figure 3 of \cite{2025ApJ...994..215H}. After generating the kernels, we extract $151 \times 151$ pixel ($6 \times 6$ arcsec) cutouts centered on each galaxy and convolve them with the appropriate kernels.

\subsubsection{Construction of Photometric Data Cube}

From the PSF-matched image cutouts, we construct photometric data cubes for the spatially resolved SED fitting. We utilize \texttt{piXedfit} to generate photometric data cubes. First, we define the region of interest for each galaxy by creating segmentation maps in all filters using \texttt{SEP} with consistent parameters: a detection threshold (\texttt{thresh} = 1.5), which selects pixels $\geq1.5\sigma$ above the background; deblending threshold (\texttt{deblend\_nthresh} = 25), controlling how finely overlapping sources are separated; and the minimum contrast ratio (\texttt{deblend\_cont} = 0.0095), allowing components with $\geq0.95\%$ of the parent flux to be deblended as separate sources. The final segmentation map is taken as the union of detections in the three longest-wavelength bands, which best trace the stellar mass distribution. After defining the galaxy's region, \pixedfit\ calculates the flux for each pixel within that area. This process is performed across all filters, generating a photometric data cube.

\subsubsection{Spatial Pixel Binning}
\label{sec:binning}

Next, we perform pixel binning, which is essential because individual pixel SEDs often have low S/N, making pixel-level SED fitting unreliable. Using \pixedfit\, we group neighboring pixels with similar SED shapes to achieve a target S/N threshold. The core principle of this method is to merge adjacent pixels with similar SED shapes until the specified S/N threshold is reached.

For the binning process, we apply the following criteria. We set a minimum bin diameter of 5 pixels (PSF FWHM of F444W). To ensure SED similarity within each bin, we impose a reduced \(\chi^2\) limit of 5. We set a signal-to-noise (S/N) threshold of 5 for the NIRCam filters and 0 for all ACS filters, reflecting the generally lower S/N in HST images compared to JWST. Additionally, for all filters probing rest-frame wavelengths shorter than the 4000\AA\ break, we assign an S/N of 0, as quenched galaxies are expected to be very faint in this regime. This process results in {51,867 bins}.

\begin{figure*}
    \centering
    \includegraphics[width=\linewidth]{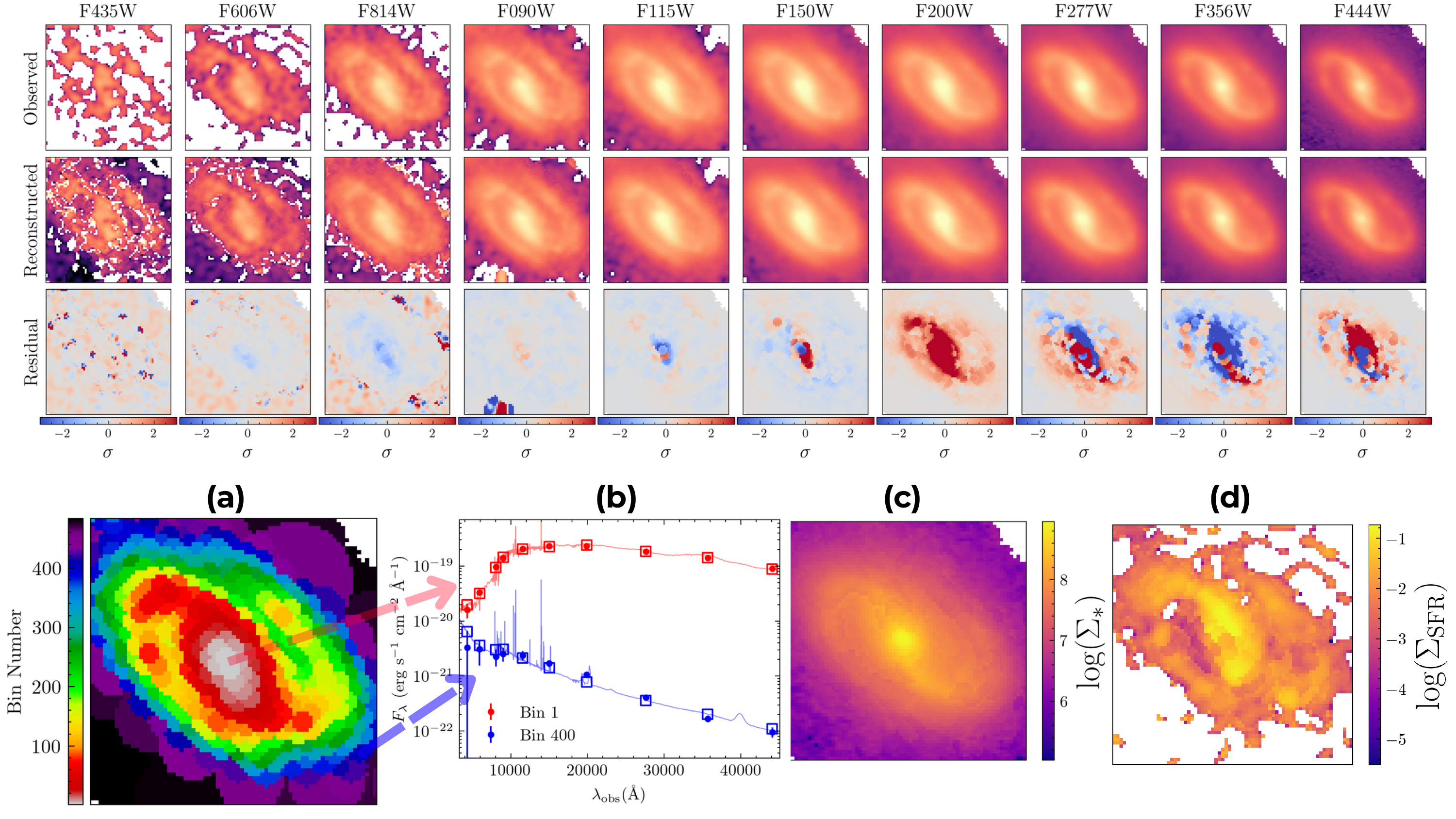}
    \caption{Example of spatially resolved SED fitting results for primer-cosmos-west-grizli-v7.0\#35188 ($z_{\mathrm{phot}}\sim1.13$). \textbf{Upper panels}: the observed multi-band images (top), reconstructed images (middle), and the residual (bottom). \textbf{Lower panels}: (a) binned image, with each bin showing different color; (b) SED examples for bin\#1 and bin\#400, with filled circles and hollow squares each showing the observed and fitted fluxes; (c) the stellar mass surface density map from the fitting, with unit of $M_\odot$ kpc$^{-2}$; (d) the star-formation rate surface density map from the fitting, with unit of $M_\odot$ yr$^{-1}$ kpc$^{-2}$.}
    \label{fig:fitting_example}
\end{figure*}

Examples of the binned images produced by \pixedfit\ are shown in Figure \ref{fig:fitting_example} (panel a). Each color represents a different bin, ranging from reddish tones in the central regions to purplish hues in the outskirts.

\subsubsection{Stellar Population Modelling} \label{sec:SED_fitting}

\begin{table*}
    \centering
    \caption{Assumed priors in the SED modelling}
    \label{tab:fitting}
    \begin{tabular}{l l l l}
    \hline
        Parameter & Description & Prior & Sampling \\
        \hline
        $\log(M_*)$ & Stellar mass & Uniform: min $=\log(s_{best})-2$, max $=\log(s_{best})+2$ & Logarithmic \\
        
        $\log (Z_*/Z_\odot)$ & Stellar metallicity & Uniform: min $=-1.0$, max $=0.2$ & Logarithmic \\
        
        $\log(t)$ & Time since star formation started & Uniform: min $=0.1$ Gyr, max = maximum age of & Logarithmic \\
         & & the universe in the templates & \\

        $\log(\tau)$ & Star-formation  decaying timescale & Uniform: min $=-1$, max $=1.5$ & Logarithmic \\

        $\hat{\tau}$ & Dust optical depth & Uniform: min $=0.0$, max $=3.0$ & Linear \\
        & in \cite{2000ApJ...533..682C} & & \\
        \hline
    \end{tabular}
    \raggedright
    NOTE: $s_{best}$ is the normalization of model SED derived from the initial fitting with the $\chi^2$ minimization method (see Section 4.2.1 in \citealt{2021ApJS..254...15A}).
\end{table*}

\begin{figure*}
    \centering
    \includegraphics[width=\linewidth]{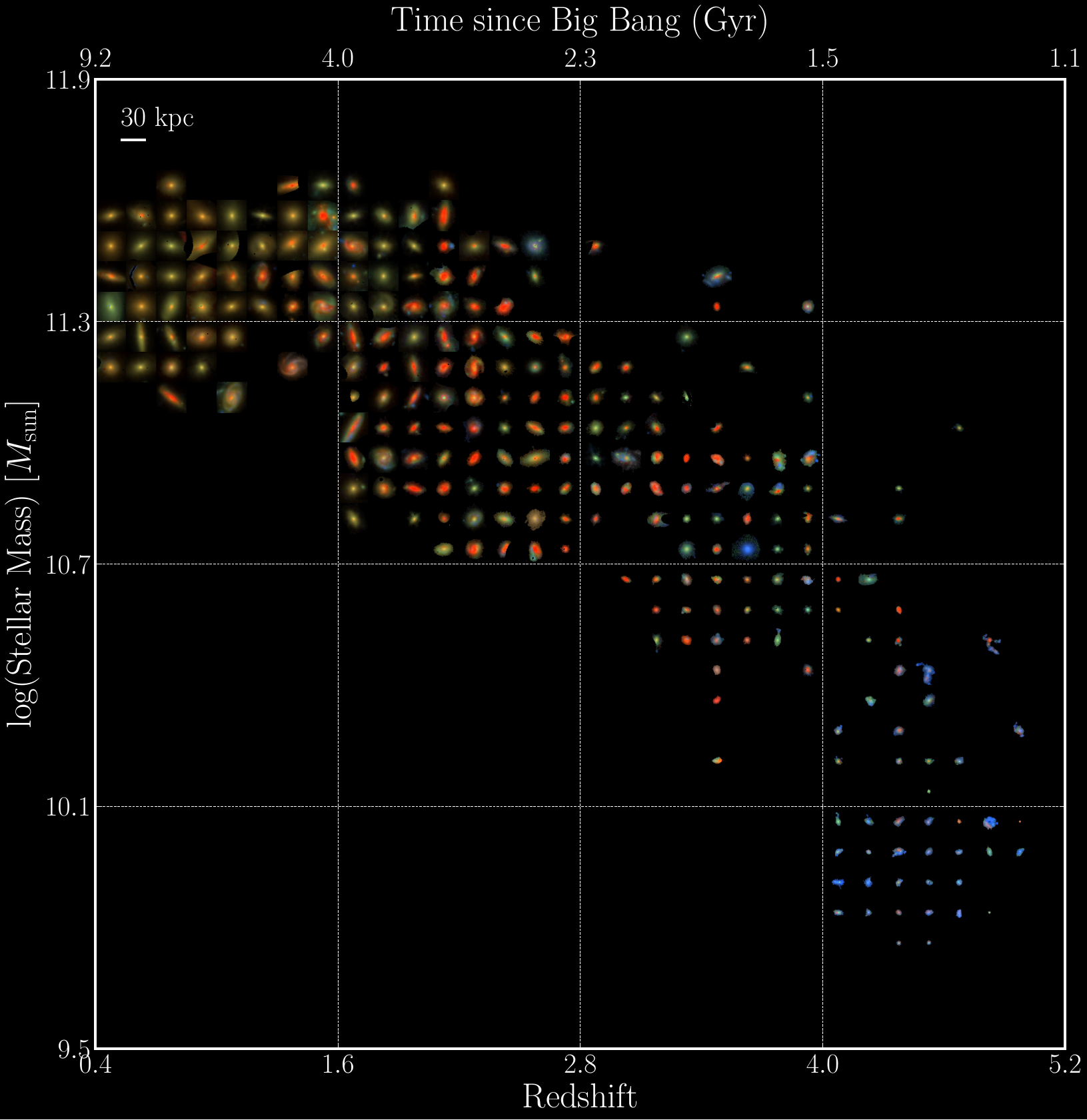}
    \caption{Pseudo three-color (RGB) images of the galaxies in our sample. The blue, green, and red channels correspond to the rest-frame U, V, and J bands, reconstructed from spatially resolved SED fitting using \texttt{piXedfit}, with the AGN nuclear component subtracted for 150/250 sources. The positions of the galaxies in this diagram reflect their redshift and stellar mass. Each galaxy image is cropped to a fixed comoving size of 30 kpc × 30 kpc.}
    \label{fig:most_massive_uvj_images}
\end{figure*}

We perform spatially resolved SED fitting on the binned data cubes using \pixedfit, which employs a fully Bayesian approach to infer stellar population properties in each spatial bin.

\pixedfit\ utilizes the Flexible Stellar Population Synthesis code (\texttt{FSPS}; \citealt{2009ApJ...699..486C}, \citealt{2010ascl.soft10043C}), which includes nebular emission modeling \citep{2017ApJ...840...44B} based on the \texttt{CLOUDY} code \citep{1998PASP..110..761F, 2013RMxAA..49..137F}. In our analysis, we adopt the initial mass function (IMF) from \cite{2003PASP..115..763C}, Padova isochrones \citep{2000A&AS..141..371G, 2007ASPC..374...33M, 2008A&A...482..883M}, and the MILES stellar spectral library \citep{2006MNRAS.371..703S, 2011A&A...532A..95F}. We model the star formation history in the form of the delayed-tau, parameterized as
\begin{equation}
    \text{SFR}(t)\propto t e^{-t/\tau}.
    \label{eq:delayedtau}
\end{equation}
Dust attenuation is modeled using the attenuation law from \cite{2000ApJ...533..682C}. Since our dataset does not include mid-infrared or far-infrared photometry and is expected to be free from AGN contamination (see Section \ref{subsec:byebyeAGN}), we switch off the dust emission and AGN dusty torus emission models. Additionally, we fix the ionization parameter to be $U=0.01$ in the nebular emission modeling. We note that this assumption may affect the inferred SFRs at the $0.1$–0.2 dex level, particularly at high redshift in strongly star-forming regions.

A summary of the adopted parameters and priors is given in Table~\ref{tab:fitting}. To accommodate the wide range of possible galaxy ages, we construct separate template libraries for different redshift intervals, each with a width of $\Delta z = 0.5$ (10 libraries in total). Each library contains 20,000 templates, yielding 200,000 templates overall. We perform the SED fitting using the random dense sampling of parameter space (RDSPS) method, which is computationally more efficient than Markov Chain Monte Carlo (MCMC) while still providing robust parameter estimates \citep{2021ApJS..254...15A}. The fitting is carried out independently for each spatial bin in each galaxy.

Once the fitting has been performed for all spatial bins in a galaxy, maps of stellar population properties can then be constructed. We only focus on the maps of stellar mass and SFR in this work. A unique feature of \pixedfit\ is its ability to recover the original spatial resolution of the input images after performing SED fitting on the binned data. Specifically, the total stellar mass and SFR derived for each bin are redistributed back to individual pixels within that bin. This redistribution is guided by the rest-frame red-band flux for stellar mass (F444W) and blue-band flux for SFR (the band with rest-wavelength closest to 2000 \AA), preserving the spatial structure traced by the observed photometry. As a result, the final stellar mass and SFR maps retain the same sampling and resolution as the original fluxmaps.

Another output of the fitting process is the set of best-fit spectra for each spatial bin, which allows us to construct a spectral data cube for each galaxy. This also enables us to reconstruct galaxy images at arbitrary bands or wavelengths. The top three rows of Figure~\ref{fig:fitting_example} show, for a galaxy at $z_{\mathrm{phot}} \sim 1.13$, the observed multi-band images; the reconstructed images in the observed JWST and HST filters, where the bin fluxes are redistributed to individual pixels following the observed distribution; and the residual images (observed minus reconstructed, normalized by the observational uncertainty).

The lower panels of Figure \ref{fig:fitting_example} presents examples of the fitted SEDs (panel b), along with the resulting parameter maps that include stellar mass surface density ($\Sigma_*$, panel c) and SFR surface density ($\Sigma_{\mathrm{SFR}}$, panel d). In the fitted SEDs, open squares represent the observed fluxes, while solid circles indicate the fluxes predicted by the best-fit models. 

In Figure \ref{fig:most_massive_uvj_images}, we present RGB images of the galaxies in our sample, where the blue, green, and red channels correspond to the rest-frame U, V, and J bands, reconstructed from spatially resolved SED fitting. Each galaxy is positioned in the diagram according to its redshift and stellar mass, consistent with Figure \ref{fig:massevol}. From Figure \ref{fig:most_massive_uvj_images}, we can see a trend where the higher redshift galaxies are bluer and smaller in size, whereas the lower redshift galaxies are redder and more extended.

\subsection{Estimation of Black Hole Accretion Rates}
\label{sec:BHAR_estimation}

\subsubsection{X-ray Luminosities of Individually Detected Sources}

In this work, we parameterize AGN activity using black hole accretion rates, which are derived from X-ray luminosities inferred from the observed X-ray count rates. For X-ray-detected sources, we adopt the hard-band count rates from the catalogs described in Section~\ref{subsec:QSOmatch}. If a source is not detected in the hard band, we instead use the full-band count rate; if it is also not detected in the full band, we use the soft-band count rate. We adopt the cataloged count-rate uncertainty when available. When no uncertainty is reported, we assume an uncertainty of $0.25\times \mathrm{CR}$, following \cite{2020A&A...642A..65C}.

\begin{figure}
    \centering
    \includegraphics[width=\linewidth]{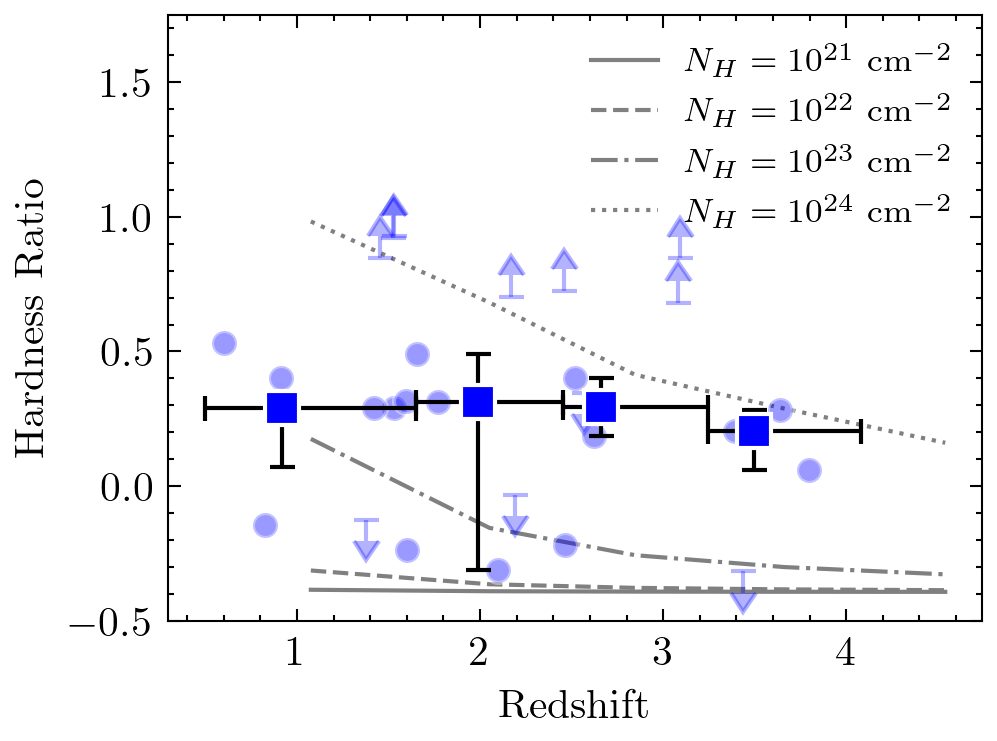}
    \caption{Blue circles show measured hardness ratio values for individually X-ray–detected sources. Sources detected only in the hard (soft) band are shown as lower (upper) limits. Blue squares indicate the median values in each redshift bin, calculated using the Kaplan–Meier estimator to properly account for censored data (upper and lower limits). The uncertainty of the HR medians are calculated via bootstrap resampling by repeatedly recomputing the median from resampled datasets and taking the 16th–84th percentile range. Grey lines show the predicted hardness ratios for different hydrogen column densities in COSMOS field.}
    \label{fig:HR_values}
\end{figure}

\begin{figure*}
    \centering
    \includegraphics[width=\linewidth]{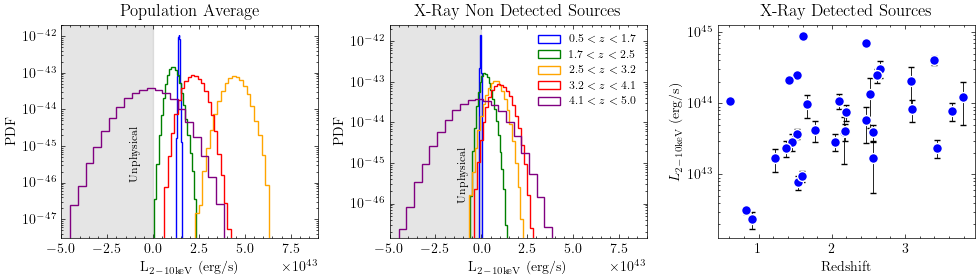}
\caption{From left to right: probability distributions of the population-averaged luminosities in each redshift bin, obtained by combining the stacked fluxes of X-ray-undetected sources with the observed fluxes of X-ray-detected sources; probability distributions of the luminosities of X-ray-undetected sources, derived from the stacked count rates; and luminosities of individually X-ray-detected sources. All luminosities are rest-frame 2--10 keV luminosities and have been corrected for galactic and intrinsic absorption, as well as for the stellar X-ray contribution from X-ray binaries.}
    \label{fig:Lxray_dist}
\end{figure*}

Because these sources are expected to be affected by intrinsic X-ray absorption, we correct for this effect by estimating the hydrogen column density required to reproduce their observed hardness ratios (HRs). The hardness ratio is defined as
\begin{equation}
\mathrm{HR} = \frac{H - S}{H + S},
\end{equation}
where $H$ and $S$ are the count rates in the hard and soft bands, respectively. For sources detected in both bands, we directly measure the HR (shown as blue circles in Figure \ref{fig:HR_values}). For sources detected in only one band, the count-rate uncertainty in the non-detected band is treated as an upper limit. Therefore, sources detected only in the hard band provide lower limits on HR, while sources detected only in the soft band provide upper limits, as shown in Figure~\ref{fig:HR_values}.

To estimate the hydrogen column density, we compare the observed HRs with model predictions, following \citet{2022ApJ...929...53I}. The model HRs are computed from absorbed power-law spectra generated with PIMMS\footnote{\url{https://cxc.harvard.edu/toolkit/pimms.jsp}} \citep{1993Legac...3...21M}, using the appropriate auxiliary response files for the Cycle 9, 14, and 16 observations of the EGS, COSMOS, and UDS fields, respectively. We fix the photon index to $\Gamma = 1.8$ and vary the intrinsic column density logarithmically over the range $10^{21}$--$10^{24}\mathrm{cm}^{-2}$ in steps of 0.1 dex. Galactic absorption is fixed to $N_\mathrm{H} = 2.6 \times 10^{20}\mathrm{cm}^{-2}$ \citep{2005A&A...440..775K}. Some examples of the model HRs are overlaid on the observed measurements in Figure~\ref{fig:HR_values}. For sources detected in both the hard and soft bands, the best-fit $N_\mathrm{H}$ is determined by minimizing the difference between the observed and model HRs.

For sources detected in only one band, the individual HRs are poorly constrained. We therefore assign to these sources the median HR in the corresponding redshift bin. The median HR is estimated using the Kaplan--Meier estimator, which accounts for both detections and censored data, including upper and lower limits. Its uncertainty is derived through bootstrap resampling, by repeatedly recomputing the median from resampled datasets and taking the 16th--84th percentile range. These median HR values are then used to estimate the corresponding hydrogen column densities for sources detected in only one band.

Finally, we use the best-fit $N_\mathrm{H}$ values and PIMMS to convert the count rates of individually detected sources into absorption-corrected 2--10 keV fluxes. We then convert these fluxes into rest-frame 2--10 keV luminosities using the luminosity distance corresponding to each source redshift, assuming the same power-law spectral shape. To quantify the luminosity uncertainty, since the conversion from count rate to luminosity is linear, we propagate the count-rate uncertainty by applying the same conversion factor to the count-rate uncertainty.

\subsubsection{X-ray Stacking of Individually Undetected Sources}
\label{subsec:stackstack}

To estimate X-ray luminosities for sources that are not individually detected by Chandra, we perform X-ray stacking using the CSTACK v4.5 tool\footnote{\url{https://lambic.astrosen.unam.mx/cstack/}} \citep{2008HEAD...10.0401M}. We focus on the hard-band count rates, measured in the observed 2--8 keV band, which are obtained by averaging the background-subtracted count rates at the input positions of the stacked sources. CSTACK provides bootstrap resampling results based on 500 realizations of the stacked count rates. We apply this stacking procedure independently for each field and redshift bin, resulting in a stacked count-rate probability distribution for each redshift bin.

Following the procedure described in the previous section, we use the median $N_\mathrm{H}$ value in each redshift bin, together with PIMMS, to convert the stacked count-rate distributions into absorption-corrected 2--10 keV flux distributions. Finally, we convert these flux distributions into rest-frame 2--10 keV luminosity distributions using the luminosity distance corresponding to the midpoint of each redshift bin, assuming a power-law spectrum with $\Gamma = 1.8$. This yields the rest-frame 2--10 keV luminosity probability distribution for the non-detected sources in each redshift bin. 

\subsubsection{Population-averaged X-ray Luminosities}

To quantify the average AGN activity in each redshift bin, we combine the X-ray fluxes from individually detected sources with the stacked signals from non-detected sources. We construct the probability distribution of the average X-ray flux in each redshift bin by randomly sampling the flux distributions of both components. For the stacked signals, we use the flux probability distributions obtained from the stacking analysis. For individually detected sources, we assume that the flux distribution of each source follows a Gaussian distribution centered on the observed flux, with the corresponding flux uncertainty as the standard deviation.

Following \cite{2015ApJ...800L..10R} and \cite{2020A&A...642A..65C}, we compute the average flux in each redshift bin as
\begin{equation}
    F_{\text{ave}}=\frac{\Sigma_{i=1}^{n_{\text{field}}}\left(\Sigma_{j=1}^{n_{i,\text{detected}}} F_{i,j,\text{detected}}+n_{i,\text{stacked}}F_{i,\text{stacked}}\right)}{\Sigma_{i=1}^{n_{\text{field}}}\left(n_{i, \text{detected}}+n_{i, \text{stacked}}\right)}
\end{equation}
where $n_{\rm field}$ is the number of fields, $n_{i,{\rm detected}}$ is the number of X-ray–detected sources in field $i$, $F_{i,j,{\rm detected}}$ is the flux of each detected source, $n_{i,{\rm stacked}}$ is the number of stacked sources in field $i$, and $F_{i,{\rm stacked}}$ is the stacked flux for that field. To estimate the uncertainty, we randomly sample the flux distributions of both the detected and stacked sources 100,000 times and recompute $F_{\rm ave}$ in each realization. This yields a probability distribution of the average flux in each redshift bin.

We then convert the resulting average flux distribution into an average rest-frame 2--10 keV luminosity distribution, using the luminosity distance corresponding to the midpoint of each redshift bin and assuming a power-law spectrum with $\Gamma = 1.8$.

\subsubsection{Conversion from X-ray Luminosity to BHAR}
\label{subsec:bhar}

From the previous sections, we obtain absorption-corrected rest-frame 2--10 keV luminosities and their uncertainties for individually X-ray-detected sources, as well as luminosity probability distributions for X-ray non-detected sources and for the population-averaged luminosity in each redshift bin. In this section, we convert these X-ray luminosities into black hole accretion rates.

Before performing this conversion, we correct for possible contamination from stellar X-ray emission within the host galaxies. To estimate the contribution from stellar processes to the rest-frame 2--10 keV emission, we adopt the empirical relation from \cite{2016ApJ...825....7L}, which accounts for X-ray emission from both old and young stellar populations:
\begin{equation}
    \frac{L_{2-10\text{keV}}}{\text{erg\: s}^{-1}}=\alpha_0(1+z)^\gamma \frac{M_*}{M_\odot}+\beta_0(1+z)^\delta \frac{\text{SFR}}{M_\odot\rm yr^{-1}},
    \label{eq:xray_bin}
\end{equation}
where $\log(\alpha_0)=29.37$, $\log(\beta_0)=39.28$, $\gamma=2.03$, and $\delta=1.31$. The first term represents emission from low-mass X-ray binaries associated with older stellar populations and scales with stellar mass, while the second term represents emission from high-mass X-ray binaries associated with young stellar populations and scales with the star formation rate.

For each individually detected source, we estimate this stellar contribution using the integrated stellar mass and SFR derived from our spatially resolved SED fitting. The corresponding uncertainty is propagated from the uncertainties in $M_\ast$ and SFR. For the luminosity distributions of non-detected sources and for the population-averaged luminosities, we estimate the stellar contribution using the median integrated stellar mass and SFR of the corresponding sample in each redshift bin. We then subtract this stellar contribution from the measured 2--10 keV luminosity to obtain the AGN-related X-ray luminosity. We show the resulting stellar-corrected luminosity distributions of the population-averaged and non-detected sources, and individually detected sources in Figure \ref{fig:Lxray_dist}.

We then convert the stellar-corrected 2--10 keV luminosities into bolometric luminosities using the luminosity-dependent bolometric correction from \cite{2020A&A...636A..73D},
\begin{equation}
K_X(L_{2-10\rm keV}) = a \left[ 1 + \left( \frac{\log(L_{2-10\rm keV}/L_\odot)}{b} \right)^c \right],
\end{equation}
where $a=15.33$, $b=11.48$, and $c=16.20$. The bolometric luminosity is then given by $L_{\rm bol} = K_X(L_X) L_X$.

Finally, we convert the bolometric luminosities into the black hole accretion rates following \cite{2012ApJ...753L..30M, 2015ApJ...800L..10R, 2020A&A...642A..65C}:
\begin{equation}
\mathrm{BHAR} = \frac{(1-\epsilon)L_{\rm bol}}{\epsilon c^2},
\end{equation}
where $c$ is the speed of light and $\epsilon$ is the radiative efficiency. We assume $\epsilon = 0.1$, which is commonly adopted for radiatively efficient AGN \citep[e.g.,][]{2020A&A...642A..65C}. The resulting accretion rates are converted into units of $M_\odot\mathrm{yr}^{-1}$. We note that some studies of X-ray-bright AGN suggest that lower radiative efficiencies may be more appropriate for certain populations, such as optically dull or elusive AGN \citep[e.g.,][]{2022IKF}.

These conversions yield BHAR values and their corresponding uncertainties for individually X-ray-detected sources, as well as BHAR probability distributions for X-ray non-detected sources and for the population-averaged measurements in each redshift bin.

\begin{figure*}[ht]
    \centering
    \includegraphics[width=\linewidth]{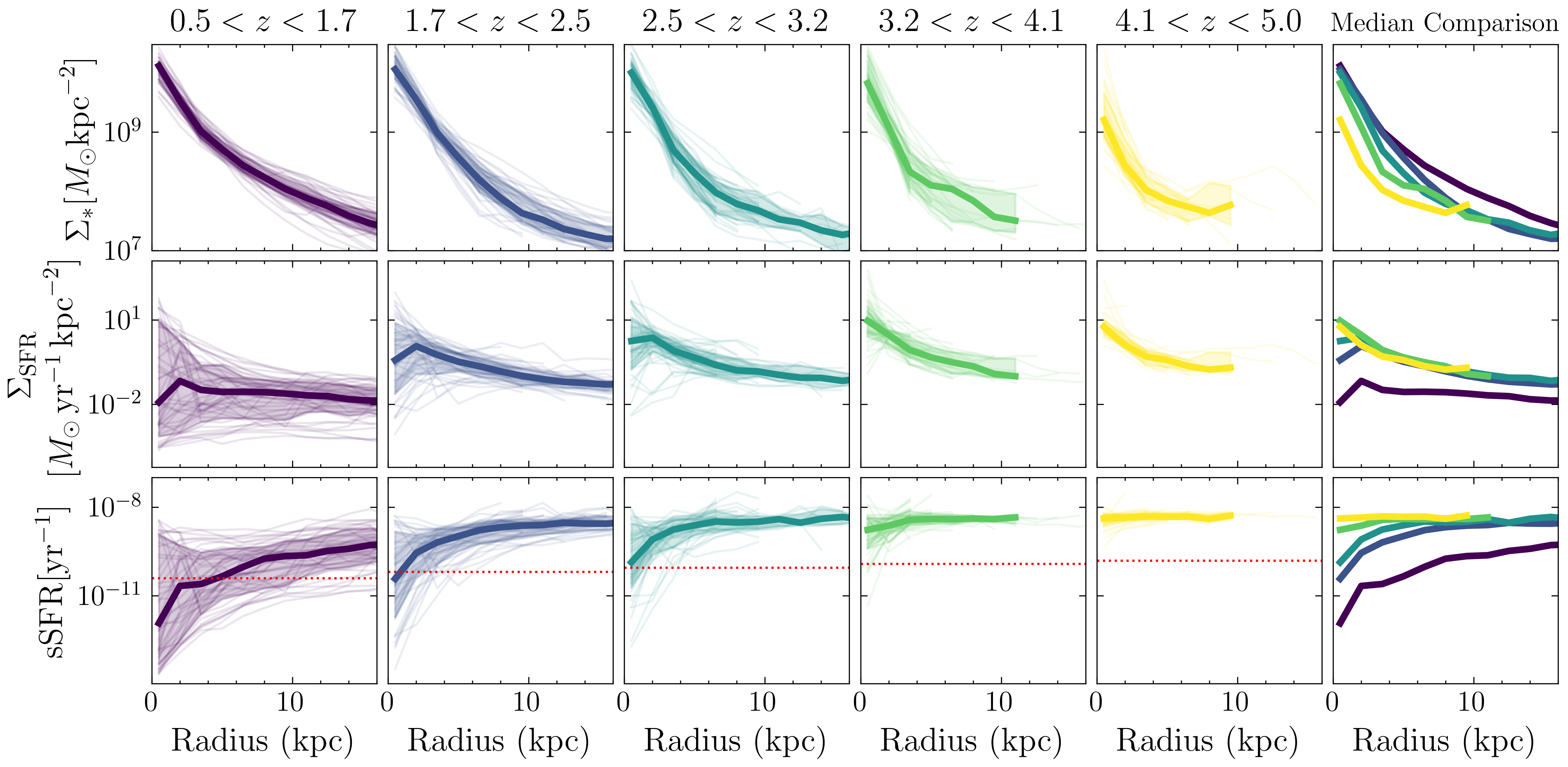}
    \caption{From top to bottom: radial profiles of stellar mass surface density, star formation rate surface density, and specific star formation rate. Solid lines and shaded regions indicate the median and the 16th–84th percentiles, respectively. Each column represents a different redshift bin, with the rightmost column showing a comparison of the median profiles across all bins. Red dotted lines in the third row indicate $0.2/t_H$, where $t_H$ is the Hubble time at the midpoint of each redshift bin.}
    \label{fig:radprof}
\end{figure*}

\begin{figure*}
    \centering
    \includegraphics[width=\linewidth]{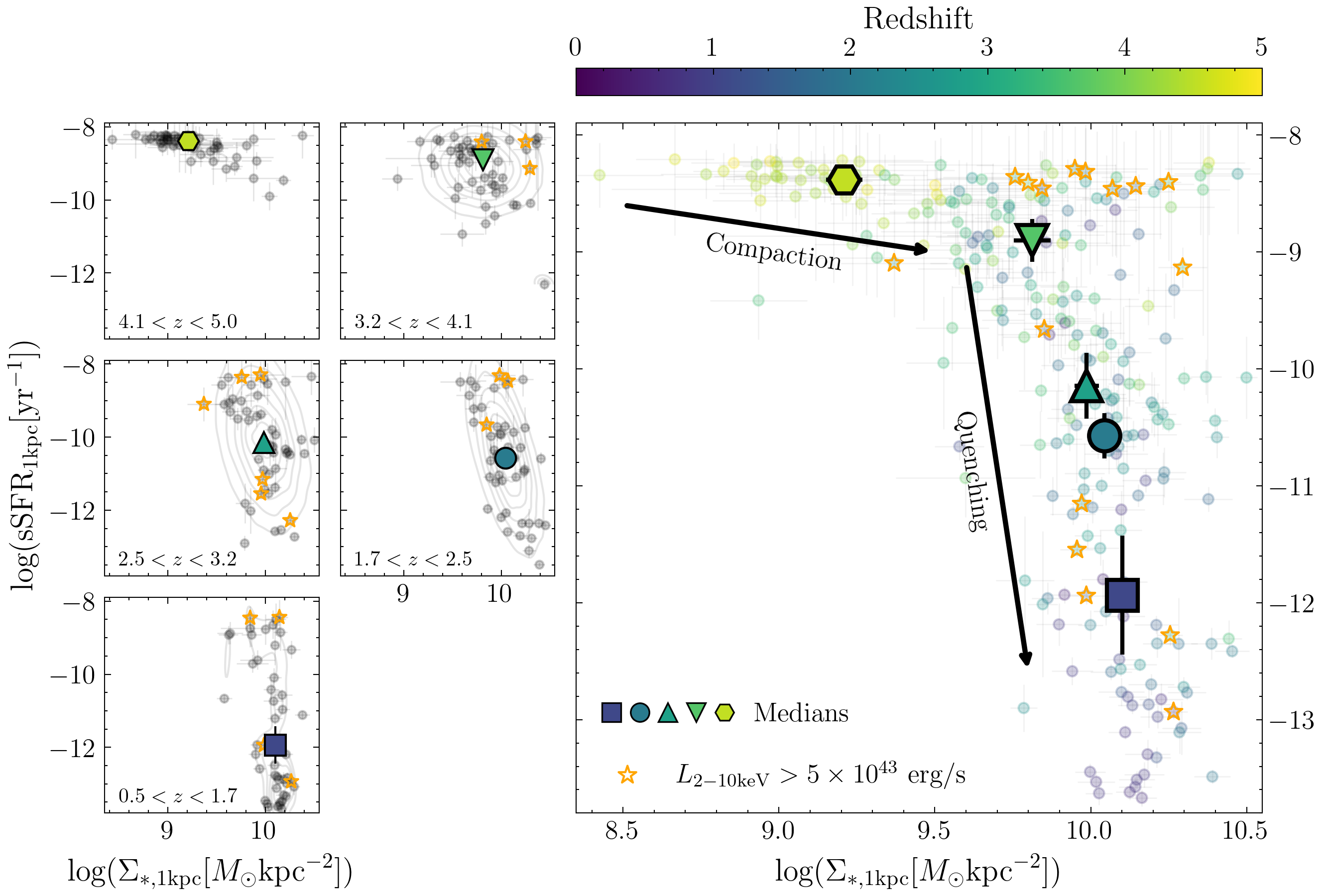}
    \caption{\textbf{Left panels}: specific star formation rate versus stellar mass surface density within the central 1 kpc for the five redshift bins. \textbf{Grey points} represent individual galaxies, while \textbf{colored symbols} indicate the median values in each bin. \textbf{Star symbols} denote X-ray–luminous galaxies ($L_{2\text{–}10,\mathrm{keV}} > 5 \times 10^{43}$ erg s$^{-1}$, which is the approximate detection limit in the highest redshift bin). \textbf{Grey contours} show the best-fit GMM distributions for each redshift bin (see Section~\ref{subsec:coregrowth}). \textbf{Right panel}: combined distribution of central (1 kpc) specific star formation rate versus stellar mass surface density, color-coded by redshift.}
    \label{fig:sSFR_M1kpc_evol}
\end{figure*}

\section{Results}

Here, we present the results of our spatially resolved SED fitting and X-ray stacking analysis. Throughout this section, we interpret these results under the working assumption that these most massive galaxies are evolutionarily connected across cosmic time. We note that stellar mass and SFR estimates are subject to systematic uncertainties, with median uncertainties of $\sim 0.2$ dex and $\sim 0.3$ dex for stellar mass and SFR, respectively. However, these uncertainties are not expected to qualitatively affect the trends presented in the following results.

\subsection{Radial Distributions of Stellar Mass and Star Formation}

Here we examine how stellar mass and star formation activity are distributed spatially in the most massive galaxies across cosmic time. Figure~\ref{fig:radprof} shows the radial profiles of stellar mass surface density, star formation rate surface density, and specific star formation rate (sSFR) (from top to bottom) in different redshift bins. Solid lines and shaded regions indicate the median profiles and the 16 to 84 percentile distribution.

The stellar mass profiles reveal a clear two-phase growth pattern. At high redshift ($z>4$), the mass distribution relatively has flatter problem. From $z>4$ to $z<4$, the central regions experience rapid growth, building up a dense core. The central stellar mass surface density increases continuously until it reaches $\sim10^{10}M_\odot\mathrm{kpc}^{-2}$ by $z\sim3$, after which the growth significantly slows. In contrast, the outer regions continue to grow toward lower redshift, indicating on-going mass assembly in the outskirts.

The evolution of star formation mirrors this structural growth. At $z>3$, the sSFR profiles are nearly flat, indicating that star formation is spatially uniform across the galaxy. At $z<3$, however, star formation in the central regions begins to decline, producing centrally suppressed sSFR profiles. This suppression becomes progressively stronger at lower redshifts and extends to larger radii, a signature of inside-out quenching. For reference, the red dotted lines in each redshift bin indicate the commonly used threshold of quenched star formation $0.2/t_H$ \citep[e.g.,][]{10.1093/mnras/sty2169}, where $t_H$ is the Hubble time at the center of the redshift bin.

\subsection{Central Stellar Density and Star Formation Across Cosmic Time}
\label{subsec:coregrowth}

\begin{figure}
    \centering
    \includegraphics[width=\linewidth]{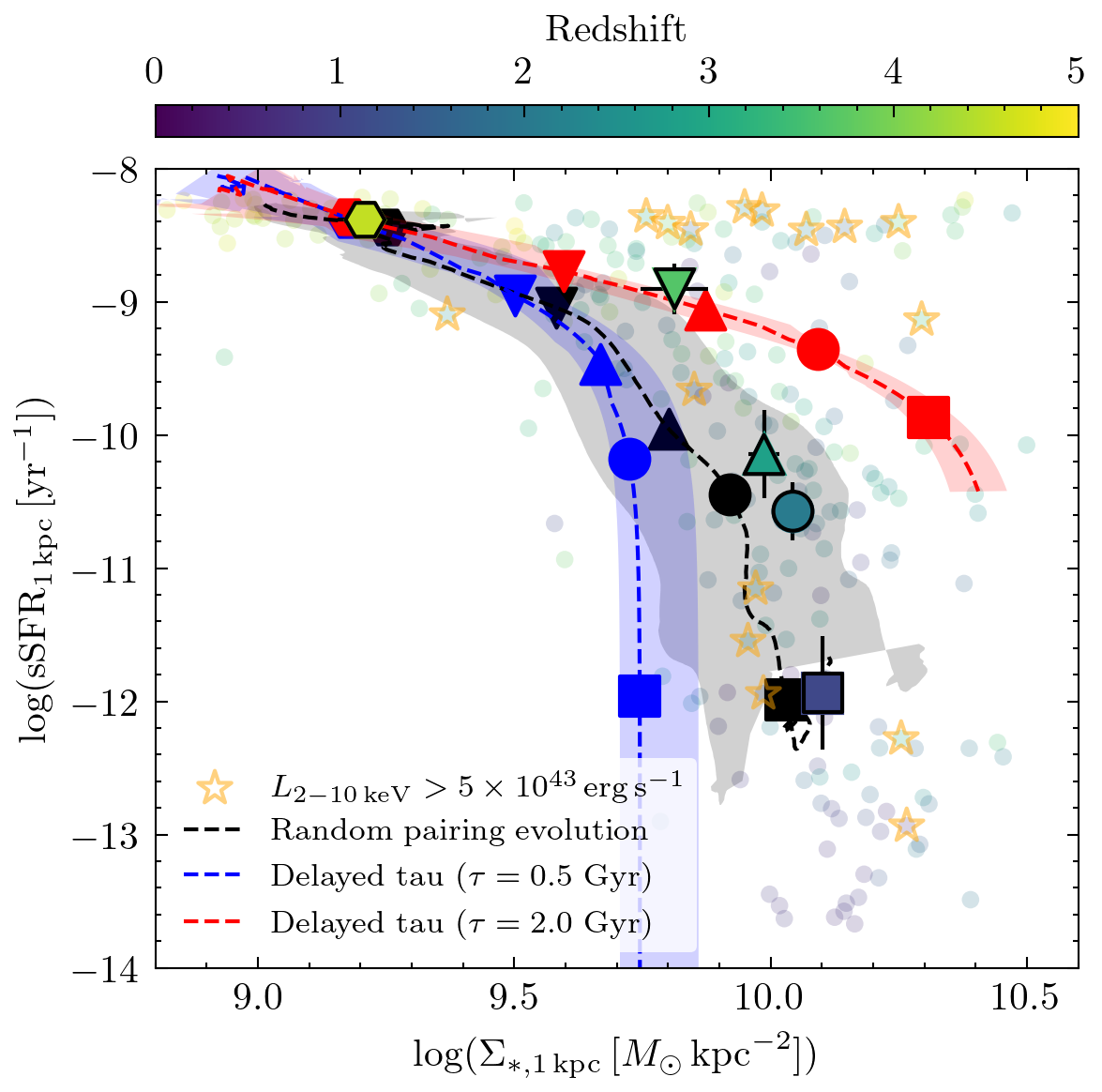}
    \caption{\textbf{Black, blue, and red curves} show the predicted evolutionary tracks from the random-pairing model and delayed-$\tau$ models with $\tau = 0.5$ and (2.0) Gyr, respectively (see Section~\ref{subsec:coregrowth}). \textbf{Shaded regions} indicate the 40th–60th percentiles of the predicted evolution, and symbols mark the expected locations for each redshift bin along the tracks. All other symbols, colors, and annotations are the same as in Figure~\ref{fig:sSFR_M1kpc_evol}.}
    \label{fig:sSFR_M1kpc_tracks}
\end{figure}

\begin{figure}
    \centering
    \includegraphics[width=\linewidth]{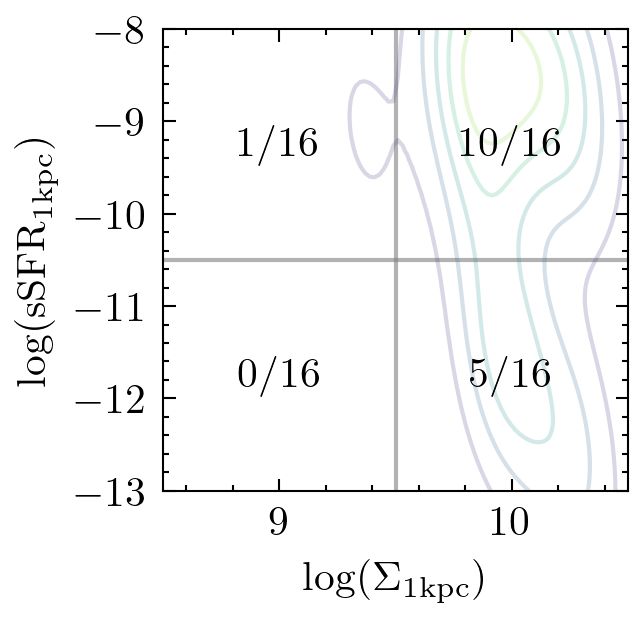}
    \caption{Fraction of X-ray–luminous sources ($L_{2\text{--}10,\mathrm{keV}} > 5 \times 10^{43}$ erg s$^{-1}$, 16 sources in total) across different regions of the $\mathrm{sSFR}_{\mathrm{1kpc}}$–$\Sigma_{\mathrm{1kpc}}$ plane. Contours indicate the underlying distribution of these sources.}
    \label{fig:xray_luminous}
\end{figure}

To investigate when and how {stellar core formation} occurs, we examine the evolution of star formation activity and stellar mass surface density within the central 1 kpc of our galaxies across cosmic time. Figure~\ref{fig:sSFR_M1kpc_evol} shows the relation between sSFR and stellar mass surface density in the {central 1 kpc} for our sample in the five redshift bins. {We use $\Sigma_{\ast,\rm 1kpc}$ as a tracer of core mass and compactness, following previous studies \citep[e.g.,][]{2015MNRAS.450.2327Z, 2016MNRAS.457.2790T, 2017ApJ...840...47B, 10.1093/mnras/stad1263}. We have also verified that our conclusions remain unchanged when adopting slightly larger aperture radii instead of 1 kpc.}

A clear evolutionary sequence emerges from high to low redshift. At $z>4$, most of the galaxies have not yet formed dense stellar cores: their central regions are diffuse and actively forming stars, occupying the upper-left part of the diagram. As time progresses from $z\sim4.5$ to $z\sim3.5$, the cores grow significantly denser ({$\sim0.60$} dex of growth) while remaining strongly star-forming, shifting toward the upper-right region. By $z\sim3$, many cores have reached sufficiently high stellar mass surface densities and their star formation begins to shut down, moving toward the lower-right region of the diagram ({$\sim1.24$} dex from $z\sim3.5$ to $z\sim2.7$). This quenching at high core density continues toward lower redshifts, with an increasing fraction of galaxies hosting compact, quenched central regions.

To assess whether the observed star formation rates alone can reproduce the observed core density trend, we construct mock evolutionary tracks based on the joint distributions of sSFR$_\mathrm{1kpc}$ and $\Sigma_{\rm 1kpc}$ in our sample. In each redshift bin, we model the distribution of galaxies in the $\mathrm{sSFR_{1kpc}}-\Sigma_{\rm 1kpc}$ plane using a Gaussian Mixture Model (GMM), which represents the data as a mixture of a finite number of Gaussian distributions.

To determine the optimal number of Gaussian components, in each redshift bin we fit the combined distribution of sSFR$_{\mathrm{1kpc}}$, $\Sigma_{\rm 1kpc}$, and redshift using GMMs with 1–11 components, and select the model that minimizes the Bayesian Information Criterion (BIC),
\begin{equation} \text{BIC}=-2\ln{{L}}+k\ln{N}, \end{equation}
where ${L}$ is the maximum likelihood, $k$ is the number of free parameters, and $N$ is the number of data points. {We use the BIC instead of AIC for the GMM selection because its stronger penalty on model complexity helps avoid overfitting the limited sample in each redshift bin}. The chosen GMM distributions for each redshift bin are shown as grey contours in Figure~\ref{fig:sSFR_M1kpc_evol}.

We then generate random pairs of galaxies by sampling from the GMM in each redshift bin and connect them across redshift by linearly interpolating their log(sSFR$_{\mathrm{1kpc}}$) and redshift, thereby constructing continuous star formation histories. To account for stochasticity in star formation, we add two oscillatory components with timescales of 0.2 and 0.5 Hubble times and amplitudes of 0.175 and 0.18 dex, respectively, following the Fourier-spectrum characterization of main-sequence variability in \cite{2016MNRAS.457.2790T}. We then integrate these star formation histories over time and add the resulting stellar mass growth to the initial $\Sigma_{\rm 1kpc}$ at the highest redshift to obtain the evolution of the core stellar mass surface density. This procedure is repeated 100,000 times. The median evolutionary track in the $\mathrm{sSFR_{1kpc}}-\Sigma_{\rm 1kpc}$ plane is shown as the black dotted line in Figure~\ref{fig:sSFR_M1kpc_tracks}.

For comparison, we also generate mock evolutionary tracks using parametric star formation histories of the delayed-$\tau$ form (equation \ref{eq:delayedtau}). In this case, we randomly sample galaxies from the highest-redshift bin GMM distribution and evolve them forward, assuming delayed-$\tau$ models with $\tau = 0.5$ and $2.0$ Gyr. Each case is again simulated 100,000 times, and the resulting median evolutionary tracks are shown in blue and red, respectively. We note that these simulations exhibit substantial scatter. For all three simulations, the shaded regions in Figure~\ref{fig:sSFR_M1kpc_tracks} indicate the 40th–60th percentile range.

We find that these simple simulations fail to reproduce the observed evolution from $4.2<z<5.0$ to $3.3<z<4.2$, as well as the large population of galaxies in the top-right region of Figure~\ref{fig:sSFR_M1kpc_evol}. Quantitatively, the random-pairing model underpredicts the median at $3.3<z<4.2$ by {0.25 dex}, while the delayed-$\tau$ models are lower by {0.33 dex} ($\tau=0.5$ Gyr) and {0.24 dex} ($\tau=2.0$ Gyr). These offsets marginally exceed the median stellar mass uncertainty of 0.2 dex. This will be discussed further in Section \ref{subsec:central_assembly_quenching}.

In Figure~\ref{fig:sSFR_M1kpc_evol}, we also highlight sources with X-ray luminosities above the approximate detection limit of the highest-redshift bin, shown as orange star symbols. {These sources tend to occupy regions of higher $\mathrm{sSFR}_{\rm 1kpc}$ and higher $\Sigma_{\ast,\rm 1kpc}$ than the rest of the population within the same redshift bin. To better quantify this trend, we divide the $\mathrm{sSFR_{1kpc}}-\Sigma_{*\mathrm{1kpc}}$ plane into four regions and compute the fraction of X-ray luminous sources in each region. The results are shown in Figure~\ref{fig:xray_luminous}, where the contours represent the underlying distribution of these sources. We find that 10/16 ({$63\%\pm12\%$}) of the X-ray luminous sources lie in the region defined by $\mathrm{sSFR_{1kpc} > 10^{-10.5}yr^{-1}}$ and $\Sigma_{*,\mathrm{1kpc}} > 10^{9.5}M_\odot\mathrm{kpc}^{-2}$.

\subsection{Relative Growth of Black Holes and Their Host Galaxies}
\label{subsec:bhar_results}

\begin{figure*}
    \centering
    \includegraphics[width=\linewidth]{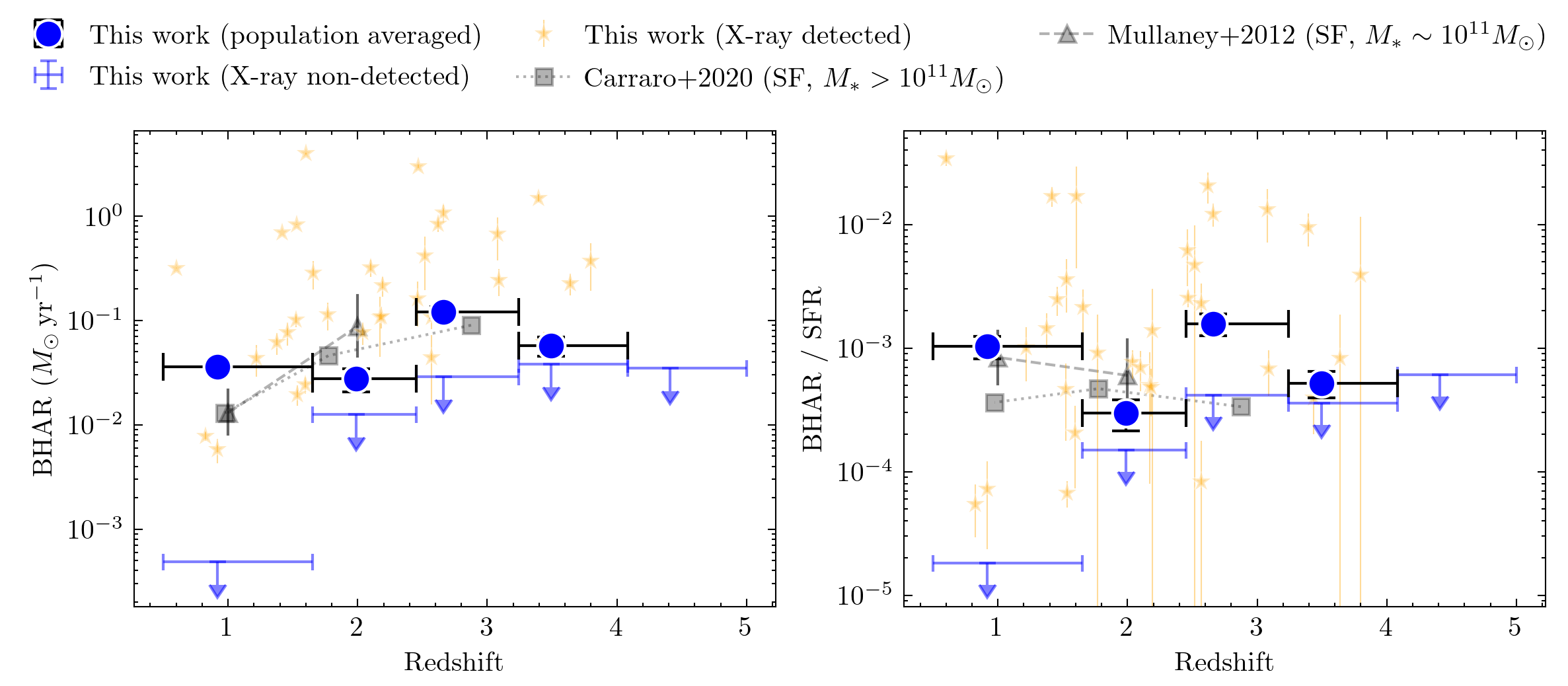}
    \caption{\textbf{Left and right panels} show the black hole accretion rate (BHAR) and the $\mathrm{BHAR}/\mathrm{SFR}$ ratio, respectively. \textbf{Blue circles, blue arrows, and orange stars} represent population-averaged values, upper limits from stacked X-ray non-detected sources, and individually X-ray-detected sources, respectively. For the population-averaged measurements, the BHAR error bars indicate the $1\sigma$ scatter of the BHAR probability distributions, while the $\mathrm{BHAR}/\mathrm{SFR}$ error bars are propagated from the $1\sigma$ scatter of the BHAR distributions and the bootstrap uncertainty of the average SFR. For the stacked signals, the \textbf{upper limits} are derived from the $1\sigma$ upper limits of the X-ray luminosity distributions. In both panels, \textbf{gray squares and triangles} show comparison measurements from \cite{2020A&A...642A..65C} and \cite{2012ApJ...753L..30M}, respectively.}
    \label{fig:BHAR_SFR}
\end{figure*}

{Having established how central stellar mass builds up and how X-ray luminous sources are distributed across the core-formation sequence, we now examine how black hole accretion evolves across cosmic time. In the left panel of Figure~\ref{fig:BHAR_SFR}, we show the average BHAR in each redshift bin, derived in Section~\ref{subsec:bhar}, as blue circles. For individually X-ray-detected sources, we show the corresponding BHAR estimates as orange stars. For X-ray non-detected sources, the stacked luminosities have low SNR; therefore, we show only their $1\sigma$ upper limits as blue arrows. We note that there are no individually X-ray-detected sources in the highest-redshift bin; therefore, in this bin, the population-averaged BHAR distribution is derived entirely from the stacked luminosity distribution.}

{We then examine how black hole accretion evolves alongside galaxy-wide star formation. To quantify the relative growth of black holes and their host galaxies, we focus on the ratio $\mathrm{BHAR}/\mathrm{SFR}$. We first calculate the average $\mathrm{BHAR}/\mathrm{SFR}$ in each redshift bin by dividing the bin-averaged BHAR obtained in Section~\ref{subsec:bhar} by the average integrated SFR of galaxies in the same bin. These values are shown as blue circles in the right panel of Figure~\ref{fig:BHAR_SFR}. The uncertainties reflect the propagated $1\sigma$ scatter in the BHAR and the $1\sigma$ bootstrap scatter of the average SFR. For X-ray non-detected sources, we estimate upper limits on $\mathrm{BHAR}/\mathrm{SFR}$ by dividing the BHAR upper limits by the average SFR of the non-detected sources in each redshift bin. These limits are shown as blue arrows in Figure~\ref{fig:BHAR_SFR}. For individually X-ray-detected sources, we compute $\mathrm{BHAR}/\mathrm{SFR}$ by dividing the BHAR of each source by its integrated SFR, and estimate the uncertainty by propagating the corresponding BHAR and SFR uncertainties. These individual measurements are shown as orange stars in Figure~\ref{fig:BHAR_SFR}.}

We find that at $z \lesssim 4$, the averaged ratios remain scattered around a roughly constant value of $\sim10^{-3}$, suggesting that the two components grow in a coupled manner. Notably, at $z \sim 3$, we observe an elevated value of {$\mathrm{BHAR/SFR} \sim 1.6\times10^{-3}$}, indicating enhanced black hole growth compared to star formation activity. At $z > 4$, our $1\sigma$ upper limit suggests that the ratio is likely below $10^{-3}$.

In Figure~\ref{fig:BHAR_SFR}, we compare our results with the massive star-forming samples of \cite{2020A&A...642A..65C} (gray squares, $\log(M_*/M_\odot)>11$) and \cite{2012ApJ...753L..30M} (gray triangles, $10.75<\log(M_*/M_\odot)<11.25$). At $z\sim3$, our BHAR values are broadly consistent with \cite{2020A&A...642A..65C}, likely reflecting a comparable stellar mass range and the dominance of actively star-forming systems at this epoch. However, our BHAR/SFR ratio at $z\sim3$ is systematically higher. This offset is likely driven by differences in SFR estimation, as \cite{2020A&A...642A..65C} derive SFRs from far-infrared stacking, which captures dust-obscured star formation and typically yields higher average SFRs compared to SED-based measurements \citep[see, e.g.,][]{10.1093/mnras/stac2291}.

At $z\sim2$, our average BHAR values fall below those of the literature samples. This is likely due to our mass-selected sample beginning to include a non-negligible fraction of quiescent galaxies, which contribute little to the average BH accretion rate and thus lower the mean BHAR. Despite this, the BHAR/SFR ratio remains comparable to the literature, suggesting that both BH growth and star formation are suppressed in a similar proportion within our sample.

At $z\sim1$, our galaxies occupy a systematically higher stellar mass regime than the comparison samples, and our selection includes a larger fraction of quiescent systems. This difference in both mass distribution and galaxy population likely drives the observed offset in BHAR relative to the literature.

\begin{figure*}
    \centering
    \includegraphics[width=\linewidth]{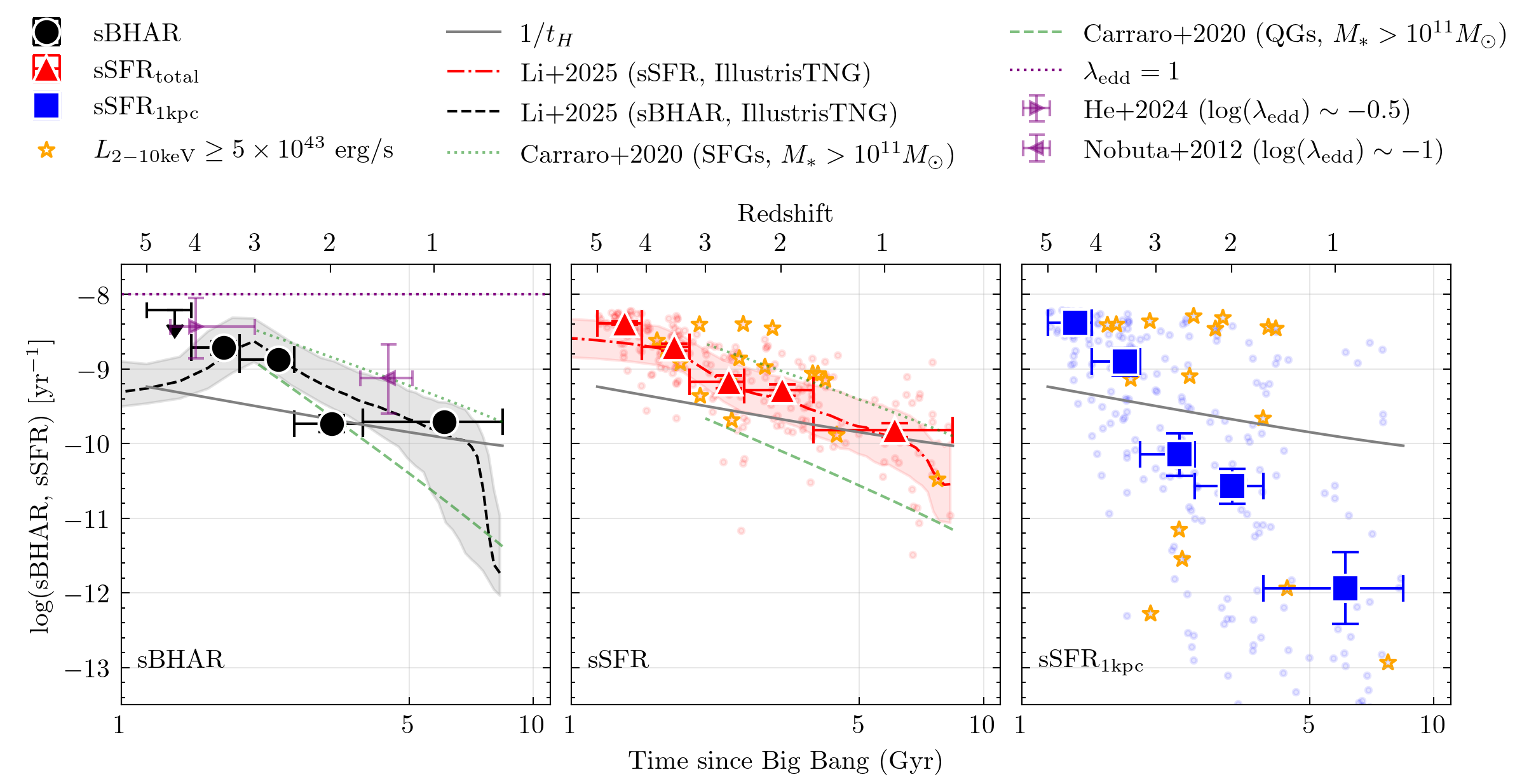}
    \caption{\textbf{Black circles, red triangles, and blue squares} show the median specific black hole accretion rate, total specific star formation rate, and central (1 kpc) specific star formation rate, respectively. Error bars on the total and central sSFR represent the $1\sigma$ bootstrapped uncertainty on the median, while sBHAR uncertainties are propagated from the BHAR scatter. The sBHAR upper limit in the highest redshift bin is derived from the BHAR upper limit and the 16th-percentile black hole mass. The \textbf{solid grey line} denotes $1/t_H$, where $t_H$ is the Hubble time. \textbf{Red dot-dashed and black dashed lines} show the sSFR and sBHAR from IllustrisTNG \citep{2025MNRAS.543.1878L}, with shaded area indicates their $1\sigma$ scatter. \textbf{Green dotted and dashed lines} indicate the measurements from \cite{2020A&A...642A..65C} for SFGs and QGs with $M_*>10^{11}M_\odot$, respectively. The \textbf{purple dotted line} shows the sBHAR for Eddington-limited growth. \textbf{Purple right- and left-facing triangles} denote Eddington ratio measurements of low-luminosity quasars from \cite{2024ApJ...962..152H} and X-ray selected broad-line AGNs from \cite{2012ApJ...761..143N}, respectively. \textbf{Orange stars} mark X-ray luminous sources ($L_{\mathrm{2-10,keV}}\gtrsim5\times10^{43}$ erg s$^{-1}$).}
    \label{fig:srate}
\end{figure*}

{Following \cite{2020A&A...642A..65C}, we use the BHAR/SFR values to estimate black hole mass growth following their host galaxy stellar mass assembly (explained more in details in Appendix \ref{app:BHmass}). }We then use these black hole mass estimates to compute the specific black hole accretion rate ($\mathrm{sBHAR}\equiv\mathrm{BHAR}/M_{\mathrm{BH}}$) and compare it with the sSFR measured within the central 1 kpc as well as the galaxy-wide value. This comparison is shown in Figure~\ref{fig:srate}. For the highest redshift bin ($z \sim 5$), we present an upper limit on the sBHAR, obtained by combining the upper limit on the BHAR with the lower bound (16th percentile) of the estimated black hole mass at that redshift. 

We find that both the total and 1 kpc sSFR decrease from $z \sim 5$ toward lower redshift, while the sBHAR shows a declining trend since at least $z \sim 4$. At $z < 4$, the sBHAR is more comparable to the total sSFR than to the 1 kpc sSFR. The 1 kpc sSFR closely follows the total sSFR at $z > 3$, but becomes significantly lower than both the total sSFR and the sBHAR at $z \lesssim 3$.

We compare our measurements with predictions from IllustrisTNG \citep{2025MNRAS.543.1878L}, shown as dashed black (sBHAR) and dot-dashed red (sSFR) lines (their Figure~3). The shown predictions adopt the median timescale (their Figure~2) and assume a transition time $t_{12}$ at $z\sim3.5$, corresponding to the median redshift of our second-highest bin where peak compaction is expected. These predictions are broadly consistent with our results, except for sBHAR at $z\sim2$, where the discrepancy reflects the relatively low stacked X-ray luminosity and BHAR (Figures~\ref{fig:Lxray_dist} and \ref{fig:BHAR_SFR}).

We further compare our results with the sBHAR and sSFR estimates from \cite{2020A&A...642A..65C} for star-forming (dotted) and quenched galaxies (dashed) with $M_* > 10^{11} M_\odot$. Based on the sSFR, our sample at $z<3$ consistently lies between their two populations. However, the sBHAR at $z\sim2$–3 is closer to their quenched galaxies, while at $z\sim1$ it approaches their star-forming sample, indicating that galaxies with similar star formation properties can exhibit different black hole accretion behavior across redshift.

Finally, we compare with Eddington ratio measurements of low-luminosity quasars at $z\sim4$ from \cite{2024ApJ...962..152H} (purple right-facing triangle), which is consistent with our results, and X-ray-selected broad-line AGNs at $z\sim1.4$ from \cite{2012ApJ...761..143N} (purple left-facing triangle), which show higher values. This difference likely reflects their selection of luminous AGNs, which preferentially trace higher accretion rates.

\section{Discussion}

\subsection{Central Stellar Mass Assembly and the Onset of Inside-Out Quenching}
\label{subsec:central_assembly_quenching}

{The evolution in the $\mathrm{sSFR}_{\rm 1kpc}-\Sigma_{\ast,\rm 1kpc}$ plane (Figure~\ref{fig:sSFR_M1kpc_evol}) shows a clear sequence in the central regions of massive galaxies. At $z>4$, most galaxies have relatively low $\Sigma_{\ast,\rm 1kpc}$ and high $\mathrm{sSFR}_{\rm 1kpc}$, indicating diffuse but actively star-forming cores. From $z\sim4.5$ to $z\sim3.5$, the median central stellar mass surface density increases dramatically by {$\sim0.60$} dex within only $\sim700$ Myr, while the central sSFR remains high. This suggests that massive galaxies build their dense stellar cores during an actively star-forming phase, rather than after the central regions have already quenched \citep[e.g.,][]{2014ApJ...791...52B, 2015MNRAS.450.2327Z, 2017ApJ...840...47B, Lapiner2023}.
}

{The same rapid central buildup is seen in the radial stellar mass profiles (Figure~\ref{fig:radprof}). At $z>4$, the stellar mass profiles are relatively extended, while at $z<4$ they become much steeper in the central region. By $z\sim3.5$, the central stellar mass surface density reaches $\sim10^{10}M_{\odot}\mathrm{kpc}^{-2}$, after which the central growth becomes slower, while the outer regions continue to grow toward lower redshift. This suggests that the inner kiloparsec is assembled earlier than the extended stellar component. This structural evolution is also visible in the reconstructed RGB images in Figure~\ref{fig:most_massive_uvj_images}: galaxies in the highest-redshift bin often appear irregular and clumpy, become most compact around $z\sim4$, and then increase in apparent size at later times, consistent with continued growth at larger radii \citep[e.g.,][]{2025ApJ...994..215H}.}

{The rapid increase in $\Sigma_{\ast,\rm 1kpc}$ is difficult to reproduce with the measured central star formation alone. In Figure~\ref{fig:sSFR_M1kpc_evol}, the mock evolutionary tracks based on the observed $\mathrm{sSFR}_{\rm 1kpc}$ distributions do not fully reproduce the transition from diffuse star-forming cores at $z>4$ to compact star-forming cores at $z\sim3.5$. This suggests that smooth or mildly stochastic star formation, as represented in these simple tracks, is not sufficient to explain the full central stellar mass buildup. The missing growth could reflect short-lived central starburst episodes that are not captured by the time-averaged SED-based SFRs \citep[e.g.,][]{2025A&A...704A.290C}, heavily obscured star formation \citep[e.g.,][]{2014PhR...541...45C, 2023ApJ...948L...8R, 2024A&A...691A.299G, 2025MNRAS.537.3453B}, or inward migration of stellar clumps \citep[e.g.,][]{2009ApJ...703..785D,10.1093/mnras/stab3810,2024ApJ...974..135J}. These possibilities are not mutually exclusive, and the present data do not require a single dominant mechanism.}

{After the dense core has formed, central star formation begins to decline. In the $\mathrm{sSFR}_{\rm 1kpc}-\Sigma_{\ast,\rm 1kpc}$ plane, this appears as a movement from the compact star-forming region toward the compact quenched region by $z\sim3$ and below, with the median $\mathrm{sSFR}_{\rm 1kpc}$ decreasing by {$\sim1.24$} dex. The radial sSFR profiles show the same behavior: at $z>3$, the sSFR profiles are relatively flat, while at $z<3$ the central sSFR becomes suppressed compared to the outskirts. The suppression also becomes stronger and extends to progressively larger radii toward lower redshift, suggesting that quenching proceeds from the inside  \citep[e.g.,][]{2025ApJ...994..215H}. Overall, the sequence from irregular and diffuse star-forming galaxies, to compact star-forming cores, and finally to centrally suppressed systems supports a picture in which massive galaxies first assemble dense stellar cores and then begin to quench from the inside out.}

\subsection{X-ray AGN Activity and Black-Hole Growth During Core Formation}
\label{subsec:discussion_xray_agn_compact_cores}

{The distribution of X-ray luminous AGN in the $\mathrm{sSFR}_{\rm 1kpc}-\Sigma_{\ast,\rm 1kpc}$ plane provides an important clue to the connection between central stellar mass assembly and black-hole growth. As shown in Figures~\ref{fig:sSFR_M1kpc_evol} and \ref{fig:xray_luminous}, X-ray luminous sources are preferentially located in systems with both high central stellar mass surface density and high central sSFR. In particular, $10/16$ ({$63\%\pm12\%$}) of the X-ray luminous sources lie at $\mathrm{sSFR}_{\rm 1kpc}>10^{-10.5}\mathrm{yr}^{-1}$ and $\Sigma_{\ast,\rm 1kpc}>10^{9.5}M_{\odot}\mathrm{kpc}^{-2}$. This suggests that luminous black-hole accretion is preferentially observed in compact star-forming cores, consistent with previous studies finding enhanced X-ray AGN incidence in compact star-forming galaxies compared to extended star-forming or compact quiescent systems \citep[e.g.,][]{2017ApJ...846..112K, 10.1093/mnras/stac2103, 2025ApJ...994..265V}.}

{This trend appears to be connected to the timing of core formation. In the highest-redshift bin, where galaxies are still relatively diffuse and their central stellar mass densities have not yet reached the compact regime, we do not detect significant black-hole accretion activity, and the population-averaged BHAR is treated as an upper limit. By contrast, at $z\sim3$--$4$, when galaxies have developed dense stellar cores while maintaining high central star formation, the incidence of X-ray luminous AGN increases and the average BHAR becomes higher. This suggests that black-hole growth becomes more observationally prominent during or shortly after the formation of compact stellar cores, consistent with previous studies finding enhanced X-ray AGN incidence in compact star-forming galaxies \citep[e.g.,][]{2017ApJ...846..112K, 10.1093/mnras/stac2103}.}

{The BHAR/SFR evolution provides a complementary view of this transition. At $z\gtrsim4$, the upper limit on BHAR/SFR likely lies below the values seen at later epochs, suggesting relatively weak black-hole growth compared to stellar growth before the compact core is established. At $z\lesssim4$, the BHAR/SFR ratios are scattered around $\sim10^{-3}$, broadly comparable to the value expected if black holes and galaxies grow in the proportion required to approach the local black-hole-to-stellar-mass \citep[e.g.,][]{2012ApJ...753L..30M, 2020A&A...642A..65C}. The emergence of this ratio around the same epoch as the rapid increase in $\Sigma_{\ast,\rm 1kpc}$ suggests a transition in the relative growth of black holes and their host galaxies during core formation \citep[][]{2021MNRAS.505..172L, 2025MNRAS.543.1878L}.}

{The comparison between sBHAR and sSFR suggests that black-hole growth is more closely related to galaxy-wide star formation than to central star formation after the onset of central quenching \citep[e.g.,][]{2026MNRAS.546ag217J}. At $z>3$, the total sSFR and central $\mathrm{sSFR}_{\rm 1kpc}$ are similar, making it difficult to distinguish whether black-hole growth is more closely connected to central or global star formation. At $z\lesssim3$, however, the central $\mathrm{sSFR}_{\rm 1kpc}$ drops below the total sSFR, indicating that the central regions are becoming quenched while star formation continues at larger radii. In this regime, the sBHAR remains more comparable to the total sSFR than to the central $\mathrm{sSFR}_{\rm 1kpc}$.}

{Overall, the absence of significant BHAR before $z\sim4$, the increased AGN incidence in compact star-forming systems, and the emergence of BHAR/SFR ratios around $\sim10^{-3}$ at $z\lesssim4$ all suggest that black-hole growth becomes more prominent around the epoch when dense stellar cores are established. After central star formation begins to decline, the continued similarity between sBHAR and total sSFR suggests that black-hole growth remains connected to galaxy-wide star-forming activity, rather than to the suppressed central star formation alone.}

\subsection{Comparison with Theoretical Expectations}
\label{subsec}

{The evolutionary sequence found in this work is broadly consistent with theoretical gas compaction models, in which dissipative processes such as gas-rich mergers, violent disk instabilities, clump migration, or low-angular-momentum gas accretion can drive gas toward the central regions, increasing the central gas density and forming compact star-forming cores \citep[e.g.,][]{2015MNRAS.450.2327Z, 2016MNRAS.457.2790T, 2017ApJ...840...47B, 2021MNRAS.505..172L, 10.1093/mnras/stad1263}. Our finding that galaxies move from diffuse star-forming cores at $z>4$ to compact star-forming cores at $z\sim3$--4, while maintaining high central sSFR, is qualitatively consistent with this picture. However, because we do not directly measure gas inflow rates, the specific physical channel responsible for the rapid central buildup remains uncertain.
}
{The subsequent decline of central sSFR is also consistent with the expected transition from compact star-forming systems to centrally suppressed systems. In theoretical models, the central gas reservoir may become depleted once gas inflow can no longer balance star formation and gas removal, allowing the inner region to quench while star formation continues at larger radii. This provides a natural explanation for the radial sSFR profiles in our sample, where central suppression appears first and then extends outward toward lower redshift. Morphological quenching may also contribute after a dense stellar core has formed, because the deeper gravitational potential can stabilize the remaining gas against fragmentation and reduce the efficiency of star formation \citep{2009ApJ...707..250M}.}

{The behavior of the X-ray AGN and BHAR/SFR evolution further resembles theoretical expectations for a transition in SMBH--galaxy co-evolution during or after compaction. Recent simulations suggest that black-hole growth can proceed from an early stage dominated by stellar growth to a later stage in which black-hole accretion becomes more efficient and more closely linked to the host galaxy \citep[e.g.,][]{2021MNRAS.505..172L, 2025MNRAS.543.1878L}. In this context, the weak or undetected BHAR at $z>4$, followed by the emergence of BHAR/SFR ratios around $\sim10^{-3}$ and an increased incidence of X-ray luminous AGN in compact star-forming cores at $z\lesssim4$, may indicate that black-hole growth becomes more prominent around the epoch of core formation.}

{The value BHAR/SFR $\sim10^{-3}$ is also interesting from an energetic perspective, because in some feedback models it is close to the level at which AGN energy injection can become comparable to, or larger than, the energy input from stellar feedback, depending on the assumed radiative and coupling efficiencies \citep[e.g.,][]{10.1093/mnras/sty1733, 2025MNRAS.543.1878L}. Thus, AGN feedback may become energetically relevant during this phase. However, this comparison is model-dependent, and our data do not directly show that AGN feedback removes or heats the central gas. We therefore interpret this result as evidence that AGN feedback is energetically plausible to drive the observed central quenching.
}
{Overall, our results are consistent with a broad theoretical picture in which massive galaxies first build dense star-forming cores, then begin to quench from the inside out, while black-hole growth becomes more prominent around the same epoch.
}


\subsection{Systematic Uncertainties and Caveats}
\label{subsec:caveats}
{
There are several caveats that should be considered when interpreting our results. First, our analysis may miss heavily obscured star formation and black-hole growth. The spatially resolved SFRs are inferred from SED fitting of rest-frame UV-to-optical emission, and therefore may not fully recover star formation that is deeply embedded in dust. If a significant amount of obscured star formation is present in the central regions \citep[e.g.,][]{2014PhR...541...45C, 2023ApJ...948L...8R, 2024A&A...691A.299G, 2025MNRAS.537.3453B}, the true central SFRs may be higher than our fiducial estimates. This could affect the inferred level of central suppression, especially during the compact star-forming and early quenching phases. Similarly, the X-ray data may miss heavily obscured or Compton-thick AGN \citep[e.g.,][]{2018ARA&A..56..625H,2025MNRAS.538.1921M}, meaning that the true incidence of black-hole growth could be higher than inferred from the X-ray detections and stacking analysis. This caveat is particularly relevant if compact star-forming cores are also dusty environments where both star formation and black-hole accretion are partially hidden.}

{Second, the accuracy of the AGN subtraction affects the inferred central stellar mass and SFR. This is particularly important because our main conclusions rely on measurements within the central kiloparsec. We discuss this issue in more detail in Appendix~A, where we perform a mock AGN injection and recovery analysis. This test shows that the expected biases are generally small, but become non-negligible in the redshift bins where AGN contamination is most relevant. In particular, we estimate that the central stellar mass is underestimated by $0.34$ dex and $0.14$ dex in the $2.45<z<3.24$ and $3.24<z<4.08$ bins, respectively. For the central SFR, we estimate an overestimate of $0.37$ dex in the $2.45<z<3.24$ bin and an underestimate of $0.15$ dex in the $3.24<z<4.08$ bin.
}
{These systematic offsets do not change our main conclusions. If anything, they would strengthen the observed trends. The underestimated stellar masses imply that the true central stellar mass buildup may be even stronger than measured in our fiducial analysis. In the $2.45<z<3.24$ bin, the overestimated SFR would also mean that the true central sSFR is lower, making the onset of central suppression stronger. In the $3.24<z<4.08$ bin, correcting the underestimated stellar mass and SFR would still support the presence of a compact star-forming core, but with an even larger central stellar mass. Therefore, uncertainties in the AGN subtraction are unlikely to drive the observed rapid central mass growth or the subsequent decline of central sSFR.}

{Another important caveat is the sample selection. We use a constant comoving number density selection to link massive galaxies across redshift, but this approach does not provide an exact progenitor-descendant mapping. Mergers, stochastic star-formation histories, and scatter in stellar mass growth can change the stellar-mass rank ordering of galaxies with time \citep[e.g.,][]{2015MNRAS.454.2770T,2017MNRAS.467.3887W,2023MNRAS.520.1774W}. As a result, a fixed number-density selection may mix different evolutionary pathways or miss some true progenitors and descendants.
}
{To assess this uncertainty, we repeat the analysis using an evolving-number-density selection following \citet{Behroozi2013}, adopting $d\log(n)/dz\simeq0.16$, where $n$ and $z$ denote cumulative number density and redshift, respectively. Setting {$n=4.4\times10^{-5}{\rm cMpc}^{-3}$} in the highest-redshift bin, this selection reduces the sample to 50, 36, 27, 20, and 14 galaxies from the highest to lowest redshift bins. Figure \ref{fig:evolnumdes} compares this selection with our fiducial constant-number-density sample, shown in red and black, respectively. The evolving-number-density selection produces systematically higher stellar masses and lower SFRs, both within 1 kpc and in the integrated quantities, as expected for a rarer, more massive population. However, the low-redshift population is not statistically distinct from the fiducial sample, while the smaller sample size weakens the population statistics. We therefore retain the constant-number-density selection as our fiducial sample and use the evolving-number-density case as a systematic test. Our main conclusions remain unchanged, although this test indicates a possible systematic shift toward higher stellar mass and lower SFR.
}
\begin{figure*}
    \centering
    \includegraphics[width=\linewidth]{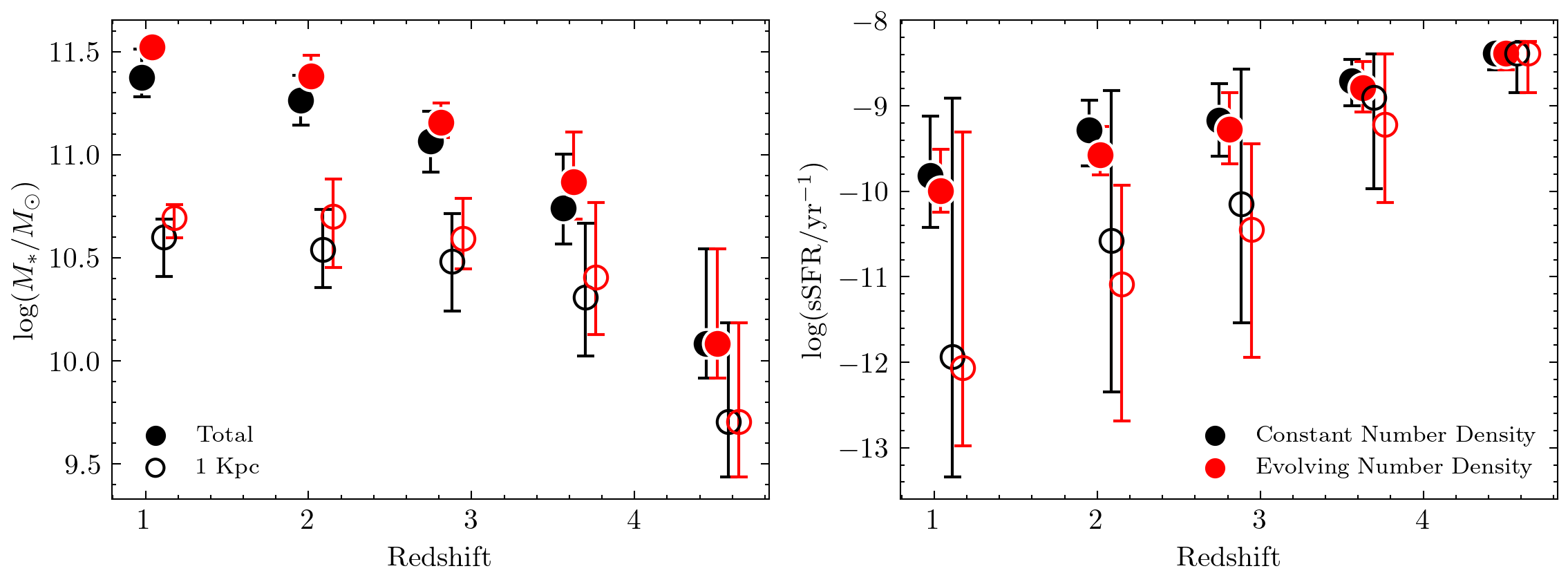}
    \caption{Comparison of stellar mass (left) and SFR (right), within 1 kpc radius (hollow circles) and integrated values (filled circles), before (black) and after (red) applying the evolving number density selection. We use the evolving number density from \cite{Behroozi2013}, assuming $d \log(n) / dz \propto 0.16$. Within each redshift bin, the data points are shifted slightly around the median redshift to reduce overlap. Error bars are showing 16th-84th percentiles.
}
    \label{fig:evolnumdes}
\end{figure*}

\section{Conclusions}

We have used spatially resolved JWST+HST imaging and Chandra X-ray data to investigate the interplay between central stellar mass assembly, star formation quenching, and black-hole growth in the number density-selected most massive galaxies at $z<5$. We account for AGN contamination using combined multi-band morphological decomposition and SED analysis, allowing us to measure central stellar mass and star formation properties more robustly. By tracing the evolution of stellar mass surface density, star formation rates on both kiloparsec and galaxy-wide scales, and average black-hole accretion, we draw the following conclusions.

\begin{enumerate}
\item \textbf{The sample selected with the constant number density exhibits evolutionary trends of rapid central stellar mass buildup at $z\sim4$.}
The central stellar mass surface density increases dramatically, with the median $\Sigma_{\ast,\rm 1kpc}$ growing by {$\sim0.60$} dex within $\sim400$ Myr. This rapid buildup is also reflected in the steepening of the radial stellar mass profiles and the increasingly compact appearance of galaxies around this epoch. Simple evolutionary tracks based on the observed central star formation rates do not fully reproduce this growth, suggesting that additional processes such as short-lived star-formation episodes, obscured star formation, mergers, clump migration, or inward redistribution of stellar mass may contribute to the assembly of dense stellar cores.

\item \textbf{Along the inferred evolutionary sequence, central star formation declines only after dense cores are established.}
Following the rapid increase in $\Sigma_{\ast,\rm 1kpc}$, the median $\mathrm{sSFR}_{\rm 1kpc}$ decreases by {$\sim1.24$} dex by $z\sim3$ and below. The radial sSFR profiles show that this decline starts in the inner kiloparsec while star formation remains elevated at larger radii. The suppression also extends to progressively larger radii toward lower redshift, indicating that quenching proceeds from the inside outward. These results suggest that dense core formation precedes the onset of central star-formation suppression.

\item \textbf{X-ray AGN activity is preferentially associated with compact star-forming cores of the sequence.}
X-ray luminous AGN are mostly found in galaxies with both high $\Sigma_{\ast,\rm 1kpc}$ and high $\mathrm{sSFR}_{\rm 1kpc}$. In particular, {$63\%\pm12\%$} of the X-ray detected sources lie in the compact star-forming region of the $\mathrm{sSFR}_{\rm 1kpc}-\Sigma_{\ast,\rm 1kpc}$ plane. In contrast, we do not detect significant black-hole accretion in the highest-redshift bin, where galaxies are still relatively diffuse. This suggests that luminous black-hole accretion becomes more common during or shortly after the formation of compact stellar cores.

\item \textbf{The relative growth of black holes and host galaxies changes around the inferred epoch of core formation.}
At $z\gtrsim4$, the upper limit on BHAR/SFR lies below the values measured at later epochs, suggesting relatively weak black-hole growth compared to stellar growth before compact cores are established. At $z\lesssim4$, the BHAR/SFR ratios are scattered around $\sim10^{-3}$, comparable to the value expected for roughly coeval black-hole and galaxy growth. The emergence of this ratio around the same epoch as the rapid increase in $\Sigma_{\ast,\rm 1kpc}$ suggests a transition in the relative growth of black holes and their host galaxies during core formation.

\item \textbf{After central quenching begins, black-hole growth remains more closely linked to galaxy-wide star formation than to central star formation.}
At $z\lesssim3$, the central $\mathrm{sSFR}_{\rm 1kpc}$ falls below the total sSFR, indicating that the inner regions are suppressed while star formation continues at larger radii. In this regime, the sBHAR remains more comparable to the total sSFR than to the central $\mathrm{sSFR}_{\rm 1kpc}$. This suggests that black-hole accretion can continue after the onset of central star-formation suppression, and may be more closely connected to the galaxy-wide star-forming activity than to the instantaneous star formation within the inner kiloparsec.

\end{enumerate}

Overall, our results reveal an evolutionary sequence in which massive galaxies first build dense, actively star-forming stellar cores, then begin to suppress star formation from the inside outward, while luminous black-hole accretion becomes more prominent around the same epoch as core formation. These trends are qualitatively consistent with theoretical scenarios linking compaction, inside-out quenching, and SMBH growth. However, the present data do not uniquely identify the physical mechanisms responsible for this sequence. Direct measurements of gas reservoirs, inflows, outflows, and feedback signatures will be needed to determine whether gas compaction, morphological stabilization, AGN feedback, or a combination of these processes drives the observed evolution.

\begin{acknowledgments}
This work is based on JWST observations obtained from the
Mikulski Archive for Space Telescopes (MAST) at the Space
Telescope Science Institute. The data used in this study are available at doi: 10.17909/f8zw-7h48. The data products
presented herein were retrieved from the Dawn JWST Archive
(DJA). DJA is an initiative of the Cosmic Dawn Center
(DAWN), which is funded by the Danish National Research
Foundation under grant DNRF140. The authors (N.S.H., J.P.
A., and R.A.S.) gratefully acknowledge the support of the
Japanese Government (Ministry of Education, Culture, Sports,
Science and Technology or MEXT) scholarship and GPPU program at Tohoku University for funding
their studies. M.A. is supported by JSPS KAKENHI Grant-in-
Aid for Scientific Research (B) grant No. 24K00670. TK acknowledges support from JSPS grant 25KJ1331. We gratefully acknowledge the funding from ITB research grant under Research and Innovation Program (PPMI) FMIPA ITB 2026 No. FMIPA.PPMI-KK-PN-08-2026

\end{acknowledgments}





%
\facilities{JWST, HST, Chandra X-Ray Observatory}

\software{astropy \citep{2013A&A...558A..33A,2018AJ....156..123A,2022ApJ...935..167A},  
          CSTACK v4.5 \citep{2008HEAD...10.0401M}, \texttt{eazy-py} \citep{2008ApJ...686.1503B, 2021zndo...7575984B}, \texttt{GALIGHT} \citep{2020ApJ...888...37D}, \texttt{nd-redshift} \citep{Behroozi2013}, 
          Source Extractor \citep{1996A&AS..117..393B}, PIMMS \citep{1993Legac...3...21M}, \texttt{piXedfit} \citep{2021ApJS..254...15A}, \texttt{PSFEx} \citep{2011ASPC..442..435B}
          }

\appendix{}
\section{Mock Recovery Tests of AGN Detection and Subtraction}
\label{app: AGN_rec}

{We perform a mock recovery analysis to quantify the uncertainty introduced by AGN identification and subtraction. In each redshift bin, we select approximately six control galaxies from our sample with no X-ray detection, $\Delta \mathrm{AIC} < 0$, and the lowest measured point-source flux fraction at rest-frame 700 nm. These galaxies were chosen to represent systems with minimal evidence for AGN contamination. For each control galaxy, we inject mock AGN point sources using 25 different combinations of dust attenuation and AGN flux fraction: five values of $E(B-V)$ and five values of $f_{\mathrm{AGN,700nm}}$. Here, $f_{\mathrm{AGN,700nm}}$ is defined as the fraction of the total flux at rest-frame 700 nm contributed by the mock AGN after injection. We then re-run the mock images through the same AGN identification and subtraction pipeline used for the real galaxies. We define a successful AGN recovery as a case where the recovered AGN flux fraction is within 0.1 dex of the injected value, the recovered $E(B-V)$ is within 0.1 mag of the injected value, and the AGN template is preferred over the stellar template, with $\Delta \mathrm{AIC} > 0$. For successfully recovered mock AGN, we subtract the recovered AGN component and then refit the central 1 kpc SED. For unsuccessful recoveries, we leave the mock AGN-injected image unchanged and fit the central 1 kpc SED directly. In both cases, we compare the recovered central stellar mass and SFR with the values measured from the original images before mock AGN injection.}

{The results are summarized in Figure \ref{fig:mock_AGN_summary}. The first row shows the AGN recovery completeness as a function of redshift, $f_{\mathrm{AGN,700nm}}$, and $E(B-V)$. The recovery rate depends primarily on AGN flux fraction rather than dust attenuation. In most redshift bins, AGN becomes detectable at $f_{\mathrm{AGN,700nm}}$ around 0.1, and the recovery is nearly complete by $f_{\mathrm{AGN,700nm}} = 0.5$. A mild dependence on $E(B-V)$ is seen at $2.5 < z < 3.2$. The low recovery rate below $f_{\mathrm{AGN,700nm}}$ around 0.1 is expected because even our control galaxies contain weak unresolved central light, with typical point-source fractions of about 0.01 -- 0.2. This unresolved component is likely dominated by compact stellar emission rather than AGN light, making very faint injected AGN difficult to distinguish from the underlying galaxy. The second and third rows of Figure \ref{fig:mock_AGN_summary} show the resulting offsets in central stellar mass and SFR. In most of the parameter space, the central 1 kpc stellar mass and SFR are recovered with small systematic offsets. The main exceptions occur in the high-$f_{\mathrm{AGN,700nm}}$ regime of the lowest-redshift bin, and in some cells at $2.5 < z < 3.2$. The low-redshift failure region does not strongly affect our real sample, because we do not find real galaxies in that part of the AGN parameter space. In contrast, at $2.5 < z < 3.2$, several real AGN occupy regions where the mock recovery shows non-negligible mass and SFR offsets. These cases therefore contribute significantly to the AGN-subtraction uncertainty in that redshift bin.}

{To quantify the expected uncertainty in each redshift bin, we combine the mock recovery results with the observed distribution of real AGN. For each bin in $f_{\mathrm{AGN,700nm}}$ and $E(B-V)$, we weight the mock-derived mass and SFR offsets by the number of real galaxies in that bin, after correcting for the mock detection completeness. The weighted mean offset gives the expected signed bias in the recovered central stellar mass and SFR. These redshift-dependent uncertainties are listed in Table \ref{tab:AGN_subtraction_errors}. Overall, the expected biases in central stellar mass and SFR are generally smaller than 0.1 dex. The main exceptions occur at $2.5 < z < 3.2$ and $3.2 < z < 4.1$. In these bins, the central stellar mass is underestimated by 0.35 dex and 0.14 dex, respectively. The central SFR is overestimated by 0.37 dex at $2.5 < z < 3.2$, but is underestimated by 0.15 dex at $3.2 < z < 4.1$. }

{Importantly, these offsets do not change our main conclusion. If anything, the negative stellar mass biases in these redshift bins suggest that the true central stellar masses may be higher than our fiducial estimates. Correcting for this effect would therefore make the inferred central mass growth stronger, rather than weaker. Thus, the mock recovery test indicates that uncertainties associated with AGN subtraction do not drive our conclusion that massive galaxies experience rapid central mass buildup.
}
\begin{figure*}
    \centering
    \includegraphics[width=\linewidth]{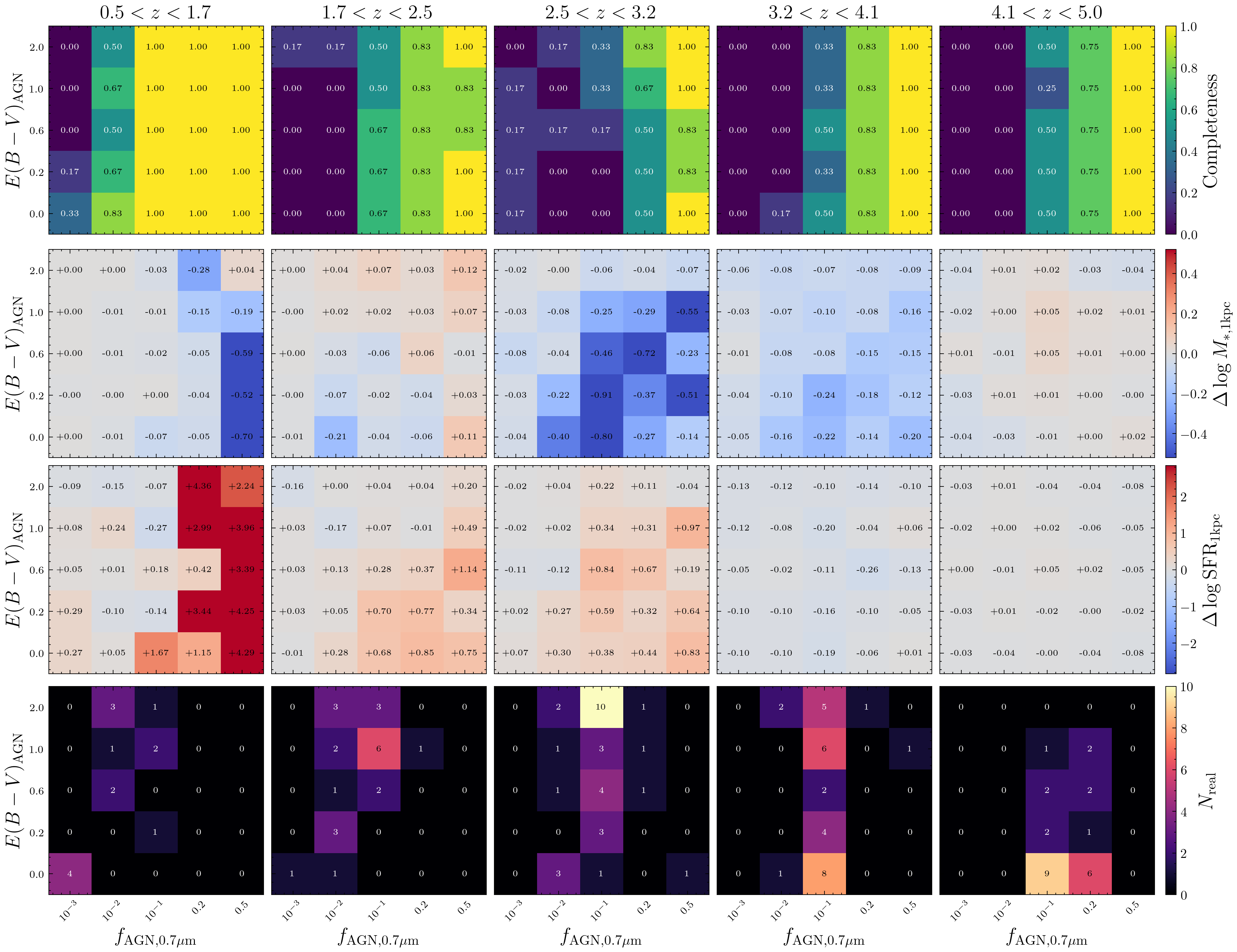}
    \caption{Top to bottom: detection completeness rate, 1kpc stellar mass offset, 1kpc SFR offset, and number of galaxies in our sample, in the $E(B-V)$ vs 7000 angstrom AGN fraction space.}
    \label{fig:mock_AGN_summary}
\end{figure*}

\begin{table}
    \centering
    \caption{The weighted average of central 1 kpc stellar mass and SFR offsets from the mock AGN point source subtraction process.}
    \label{tab:AGN_subtraction_errors}
    \begin{tabular}{l l l}
    \hline
        Redshift Range & $\Delta M_{*,\mathrm{1kpc}}$ & $\Delta \mathrm{SFR_{1kpc}}$ \\
    \hline
        $0.5 < z < 1.7$ & $-0.0028$ & $+0.071$ \\
        $1.7 < z < 2.5$ & $-0.015$ & $+0.037$ \\
        $2.5 < z < 3.2$ & $-0.34$ & $+0.37$ \\
        $3.2 < z < 4.1$ & $-0.14$ & $-0.15$ \\
        $4.1 < z < 5.0$ & $+0.01$ & $-0.0065$ \\
        \hline
    \end{tabular}
\end{table}


\section{Reconstruction of Black Hole Mass Growth}
\label{app:BHmass}

To interpret the BHAR/SFR ratio in the context of SMBH–galaxy co-evolution, we use the average BHAR/SFR ratios to reconstruct the possible growth histories of black holes and stellar masses by numerically integrating the following equation: 
\begin{equation}
    \frac{\rm BHAR}{\rm SFR} = \frac{\partial M_{\rm BH}/\partial t}{\partial M_{*}/\partial t}=\frac{\partial M_{\rm BH}}{\partial M_{*}}.
    \label{eq:BHAR_SFR}
\end{equation}

First, we construct a linear redshift grid of 1000 points spanning the centers of our lowest and highest redshift bins. For each redshift bin, we randomly sample the BHAR/SFR ratio, assuming a Gaussian distribution centered on the measured mean value with a width given by its uncertainty. For the highest-redshift bin, where only an upper limit is available, we instead sample BHAR/SFR uniformly between zero and the upper limit. These sampled values are then connected across bins and linearly interpolated as a function of the redshift grid to obtain a continuous BHAR/SFR evolution.

We follow a similar procedure for the stellar mass: in each redshift bin, values are sampled from a Gaussian centered on the median with a width given by the uncertainty of bootstrapped median, and then linearly interpolated to construct a continuous growth history.

Using Equation~\ref{eq:BHAR_SFR}, we estimate the black hole mass growth at each timestep of the redshift grid as
\begin{equation}
\Delta M_{\mathrm{BH},i}
= \int_i \frac{\mathrm{BHAR}}{\mathrm{SFR}} dM_\ast
\approx \left(\frac{\mathrm{BHAR}}{\mathrm{SFR}}\right)_i \Delta M_{\ast,i},
\label{eq:Delta_MBH}
\end{equation}

where $\Delta M_{\mathrm{BH},i}$, $(\mathrm{BHAR/SFR})_i$, and $\Delta M_{\ast,i}$ denote the black hole mass growth, BHAR/SFR ratio, and stellar mass growth at each timestep, respectively. Summing over all timesteps yields the cumulative black hole mass growth following the assembly of the stellar mass of their host galaxies. We repeat these procedures (from random sampling to Eq. \ref{eq:Delta_MBH} integration) 100,000 times to build robust statistics.

Since this method provides only the differential growth, we normalize each realization by assuming that the final black hole mass follows $M_{\mathrm{BH}} = 10^{-3} M_*$. This assumption is motivated by \cite{2012ApJ...753L..30M}, who showed that $M_{\mathrm{BH}}/M_* \approx \mathrm{BHAR/SFR}$ once significant star formation has occurred. As shown in Figure~\ref{fig:BHAR_SFR}, our measured $\mathrm{BHAR/SFR}$ values converge to a nearly constant value of $\sim10^{-3}$.

The resulting median $M_{\mathrm{BH}}$–$M_\ast$ evolutionary track is shown as a blue solid line in Figure~\ref{fig:BH_growth}, with shaded regions indicating the $1\sigma$ scatter. We caution that the absolute black hole masses should not be over-interpreted, as they cannot be robustly constrained from the X-ray data alone.

From this numerical integration, we find that the possibly lower BHAR value due to X-ray non-detections at $z > 4$ (Figure~\ref{fig:BHAR_SFR}) leads to slow black hole mass growth, resulting in predominantly horizontal evolution in the diagram. In contrast, the elevated $\mathrm{BHAR/SFR}$ at $z \sim 3$ drives rapid black hole growth, producing a more vertical evolutionary track, before ultimately converging toward the $M_{\mathrm{BH}} = 10^{-3} M_*$ relation by $z \sim 2$.

In Figure~\ref{fig:BH_growth}, we compare our results with several scaling relations from the literature. The red, yellow, and green lines show the scaling relations for local early-type galaxies (ETGs) from \cite{annurev:/content/journals/10.1146/annurev-astro-082708-101811, 2015ApJ...813...82R, 2016ApJ...817...21S}, respectively. The orange and purple dotted lines show scaling relations for AGN from \cite{2015ApJ...813...82R} and \cite{2020ApJ...889...32S}, respectively. The grey dashed and dot-dashed lines show the scaling relations from \cite{2020A&A...642A..65C} for SFGs at $z\sim2.8$ and $z\sim0.4$, respectively.
\begin{figure*}
    \centering
    \includegraphics[width=\linewidth]{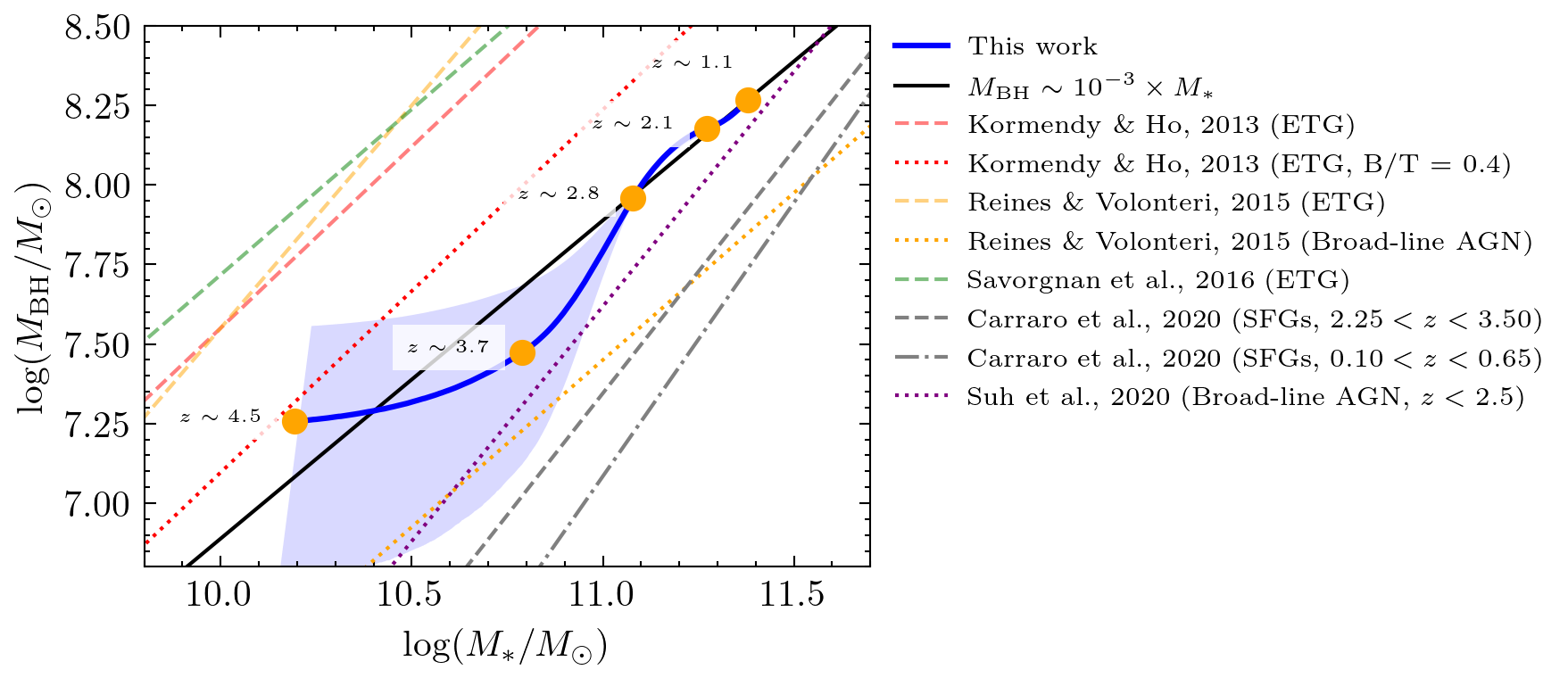}
    \caption{\textbf{Blue solid line} shows evolutionary track of black hole mass versus total stellar mass for our sample. \textbf{Yellow points} show the median values of the blue line at the midpoint of each redshift bin, while the \textbf{blue shaded region} indicates the $1\sigma$ uncertainty of the calculation. \textbf{Black line} is showing scaling relation of $M_{\mathrm{BH}}=10^{-3}M_*$. \textbf{The red dashed and dotted lines} denote the local relation from \cite{2013ARA&A..51..511K}, and the same relation with assuming bulge-to-total mass ratio of 0.4, respectively. \textbf{Orange dashed and dotted lines} show the local scaling relation of ETGs and AGNs from \cite{2015ApJ...813...82R}, respectively. \textbf{Green line} is showing the local relation for ETGs from \cite{2016ApJ...817...21S}. \textbf{Purple line} shows the scaling relation for AGN at $z<5$ from \cite{2020ApJ...889...32S}. \textbf{Grey dashed, and dot–dashed lines} show comparison relations for star-forming galaxies from \cite{2020A&A...642A..65C} at $2.25 < z < 3.50$ and $0.10 < z < 0.65$, respectively.}
    \label{fig:BH_growth}
\end{figure*}

\bibliography{references}{}
\bibliographystyle{aasjournalv7}



\end{document}